\documentclass[final,1p,times]{elsarticle}
\usepackage{setspace}
\usepackage{graphics,amssymb,amsfonts, multirow,booktabs,amsmath,hyperref,epsfig,color,eurosym, amstext,color,dcolumn,adjustbox,lscape}
\graphicspath{{figure/}}
\journal{arXiv}

\usepackage{makecell}
\usepackage[singlelinecheck=false]{caption}
\DeclareCaptionFont{tenpt}{\fontsize{10pt}{11pt}\selectfont #1}
\usepackage{float}
\hypersetup{pdfauthor={Name}}
\usepackage{subcaption}
\usepackage{endnotes}
\let\footnote=\endnote
\usepackage{natbib}
\setcitestyle{aysep={,},yysep={;}}
\usepackage{pdflscape}
\usepackage{placeins}
\usepackage{graphicx}

\begin{document}
\begin{frontmatter}
\title{The spillover effect of neighboring port on regional industrial diversification and regional economic resilience}

\author[a]{Jung-In Yeon}
\ead{yeonjungin11@gmail.com}
\author[b]{Sojung Hwang}
\ead{hwangsojung@inha.edu}
\author[b,c]{Bogang Jun \corref{cor1}}
\ead{bogang.jun@inha.ac.kr}

\cortext[cor1]{Corresponding author}

\address[a]{Sustainable Growth Initiative (SGI), Korean Chamber of Commerce and Industry, Seoul, South Korea}
\address[b]{Department of Economics, Inha University, Incheon, South Korea}
\address[c]{Department of Data Science, Inha University, Incheon, South Korea}

\begin{abstract} 
We examine the spillover effect of neighboring ports on regional industrial diversification and their economic resilience using the export data of South Korea from 2006 to 2020. First, we build two distinct product spaces of ports and port regions, and provide direct estimates of the role of neighboring ports as spillover channels spatially linked. This is in contrast to the previous literature that mainly regarded ports as transport infrastructure per se. Second, we confirm that the knowledge spillover effect from neighboring ports had a non-negligible role in sustaining regional economies during the recovery after the economic crisis but its power has weakened recently due to a loosened global value chain.
\end{abstract}
\end{frontmatter}

\newpage
\section{Introduction}
Economic development has been widely examined in various research streams in social science from the era of mercantilism to the recent globalized world. Over the history of capitalism, the detailed strategies and policies for the economic development have been evolved over time as the answer of the demand of the times. From the fall of the Iron Curtain and the ''end of history'' in the 1990s, the second wave of globalization became a global phenomenon. This new global regime changed the battleground for the economic development of firms, sub-national regions, and states. From the mid-2000s, the main drivers of the industrial transformation of a state was not the domestic factors including state-led policies, but the factors associated with complex global production networks (GPNs). With this observations, scholars in economic geography examined how regional actors and resources co-evolve with the globalized world, show the strategic coupling with plug into the GPN, and in the long run, result in the economic development of region\citep{coe2015global,yeung2016strategic, yeung2021regional}.

\cite{yeung2016strategic} explained how the industrial structure of a nation could be upgraded in the recent economic globalization and cross-border production by looking at three East Asian countries, which are Taiwan, South Korea, and Singapore. Regarding the economic development of those nations in the 1970s and the 1980s, as \cite{amsden1992asia, wade2003governing, evans1996introduction} showed, the role of state for industrial transformation was crucial and the domestic factors on the development that were associated with a legacy of Friedrich List played a critical role for the development~\citep{jun2016legacy}. However, \cite{yeung2016strategic} observed that from the 2000s, economic globalization dimmed the role of state in the industrial transformation, and transnational economic actors such as a globally significant lead firm and geographically dispersed network of overseas affiliates gave more significant effects on the industrial upgrading of a nation. He conceptualized this co-evolutionary mechanism of economic development as strategic coupling. The term ''strategic'' carries that his basic analytical significance is on firm-specific initiative, while the term ''coupling'' emphasizes that the industrial transformation should be examined as the dynamic relational process. Since \cite{yeung2016strategic} regarded firm as a basic analytical unit, this concept allowed the micro perspective in the analysis of the development. At the same time, since this GPN research ultimately examines the economic performance at the national level, it also allows the lens with macro perspective.  

However, as \cite{coe2019global} mentioned, more lens to see the interaction between the region and the GPN is required. In this regard, \cite{yeung2021regional} attempt to integrate two research streams, which are the evolutionary economic geography (EEG) and the global production networks (GPNs), by using one of their key concepts: related variety in EEG and strategic coupling in GPNs. He argued that although his endeavor of bridging the two concept directed from the GPNs to the EEG perspective, more endeavor with the opposite direction was required. 

Recent research on related variety in EEG, together with those of the economic complexity, accumulated the evidence of the role of related variety at various levels, such as a firm, a sub-national region, or a country, for various domains, such as research, product, or industry. Those empirical evidences were recently formalized as the Principle of Relatedness~\citep{hidalgo2018principle}. While in the initial work on related variety of \cite{Frenken2007}, the entropy measure the related variety, the density measure has been more used to measure the related variety following the seminar work of \cite{Hidalgo2007}. \cite{Hidalgo2007} built a product space, which is a network structure among products, by looking at co-exporting patterns of products in world trade, and calculated the density of related product for each product of each country on the product space. They found that the industrial diversification pattern exhibits path-dependency, which means that countries are more likely to enter a new product when they already export the related product to the new product. This finding holds at regional level. \cite{Boschma2013}, \cite{Gao2021}, and \cite{Jara-Figueroa2018} found that regions in Spain, China, and Brazil are more likely to enter a new industry when they already have the related industries. These findings told us that existing related industries within regions are a source of industrial diversification through inter-industry spillovers. 

However, as \cite{yeung2021regional} pointed out, the perspective of relatedness has a limitation that they only focus on local capabilities for the economic development and cannot embrace the extra-regional linkages. To mitigate the limitation, the role of inter-regional spillovers are also examined in the industrial diversification of region. \cite{bahar2014} show that countries are more likely to successfully enter a new product when they have neighboring countries that already export that product. At regional level, \cite{Gao2021} find that Chinese regions successfully diversify to a new industry when the region has neighboring regions that already have that industry. 

At the same time, examination of cross-space spillover can mitigate the limitation. Although \cite{bahar2014} and \cite{Gao2021} examine the inter-regional spillover in industrial diversification, they build one general product and industry space, respectively, and trace the product and industry trajectories on the same space, focusing on their product/industry spillover that happen only product/industry space. However, a new product/industry can emerge in a region through the spillover across different dimensions, such as between technology space and product/industry space, or skill space and product/industry space. \cite{Catalan2020} suggest the concept of cross-space spillover to describe the endogenous capabilities associated with two knowledge dimensions, which are in scientific and technological knowledge within a country. They link two dimensions by introducing a new measure, scientific-technological cross-density, and see the effect of cross-density on the technological diversification at country level. They find that the cross-density is a good predictor of countries entering a new technology. In addition to the cross-space spillover across different dimension of knowledge, cross-spillover among different types of regions can make an effect on the regions' industrial diversification.  

As the further empirical effort to bridge the strategic coupling in the GPN research and the relatedness from EEG and economic complexity, we study the role of neighboring ports, where show different types of product structure, in regions' industrial diversification and regional economic resilience, by using port-level export data and regional production data of Korea from 2006 to 2020. Port is a hub that connect the region to the GPN. When sub-national level (instead of firm level) is examined regarding the effect of GPN, it is difficult to extract the product list that is associated with the GPN. Product list that are exporting in a certain port, however, can reveal the product list plugging in the GPN and the knowledge associated with being connected to the global market or the GPNs is embedded in the products at the port. Therefore, we use the port-level export data and regional production data to examine the spillover effect of ports on the industrial diversification of neighboring regions by building two types of product-space, which are product space of ports and that of neighboring port regions, and by looking at their cross-space spillover. 

Is spillover effect from a port helpful for a region that has the port to diversify their industry? If so, between spillover from a port (source from outside of a region) and other related industries within the region (source from inside of a region), which one is the prime knowledge source of industrial diversification of a sub-national region? Do the spillover effects change during the economic crisis? 

We find that the product space of ports exhibits distinctive feature that is different from the product space of ordinary regions and countries, mainly because of containerization. Consistent with the previous research~\citep{Hidalgo2007, Neffke2011, Boschma2013, Gao2021}, a region neighboring the port is more likely to enter a new industry when it already has the related industries within their region. In addition, we find the evidence of cross-space spillover that more related products in ports increase the probability of success in neighboring regions' entering a new product. Our results imply that inter-industry spillover is the prime engine of industrial diversification in a region, and cross-space inter-regional spillover from nearby ports catalyzes the diversification of the region. In other words, the main drive of industrial diversification at sub-national level can be found within the region and their plugging into the GPNs boosts their industrial diversification. 
Finally, upon the recent change in the global trade circumstances including the trade war between China and the US, and the global pandemic crisis with COVID-19, we investigate the period-variant spillover effect of neighboring ports and discuss the economic resilience of port regions. In particular, we explore the role of the spillover effects when the regional economy faces an exogenous economic shock, such as the financial crisis in 2008. Our estimates for the years of the crisis show that strong inter-industry spillover helps port regions to sustain their industrial diversification, while the effect of cross-space spillover from neighboring ports becomes insignificant. However, during the recovery periods, cross-space spillover from nearby ports increases the probability of success in port regions’ entering a new industry, which verifies having a port nearby positively contributes to the regional resilience. Our results imply that regions' connecting the GPNs benefits their industrial diversification but for the sustainable industrial diversification, the main source should be within the regions. 

The remainder of this paper is organized as follows. Section 2 briefly reviews the recent regional studies based on relatedness and the contribution of ports to elaborate our research problems. Section 3 details the data and research methodology, especially the process of building separate product spaces, calculating two types of product relatedness, and deriving our empirical specification. Section 4 presents our results and Section 5 presents the conclusion.

\section{Literature review}
\subsection{Relatedness and knowledge spillover}
While the concept of variety and their role in economic development had been ignored in the mainstream economics, scholars in evolutionary economics have paid their attention to the role of variety in the structural change and the economic development~\citep{pasinetti1983structural, saviotti1996technological, saviotti2004economic, saviotti2020diversification}. In this stream, the degree of relatedness driving knowledge spillover has been highlighted as a determinant of regional growth paths in the field of evolutionary economic geography and regional study~\citep{Boschma2017,Hassink2019,hidalgo2018principle,Oinas2018}. \cite{Frenken2007} explained the regional economic growth by using the concept of related variety that measured by Shannon's entropy and found that the positive and significant effect of related variety on regional economic growth. 

In parallel, the literature on economic complexity has further developed this concept with empirical methodology from network science~\citep{Hidalgo2007}. As a pioneering work, \cite{Hidalgo2007} calculated proximity between products based on the probability that a country exports both products in tandem by using world trade data, suggesting the idea of the product space as a network representation of proximity. For example, the proximity between a shirt and socks are closer than that between a shirt and a car. In addition, they proposed a measure of product relatedness of each product for country as the average proximity of the product based of the country's current industry structure. For example, when a country tries to enter a car industry, the probability of success can be higher when the country already has the related product of the car, such as tire or engine industry, rather than when the country only has textile and food industry. They found that the product relatedness works as a good predictor for their future industrial structure implying the path-dependency characteristics of industrial diversification. Owing to its methodological intuitiveness and flexibility, a growing number of recent studies utilized this framework to confirm that the relatedness between industries, occupations, and patents possibly measures their localized knowledge and capabilities, and identifies with the channels of knowledge spillovers \citep{Neffke2011,Boschma2013,Kogler2013,Felipe2014,Boschma2015,Zhu2017, Jara-Figueroa2018}. The growing evidences by expanding literature, exploring the regional diversification and development as a function of the density of related activities within a region, has become formalized as empirical principle called the \textit{Principle of Relatedness}~\citep{hidalgo2018principle}.

As \cite{Boschma2016} pointed out, the studies above mentioned tend to focus more on the inter-industry spillover within a region—a kind of Jacobs externalities between industries—as if a region is a geographically self-contained entity. Of importance, however, the knowledge spillover has the spatial dimension \citep{Audretsch2004}: the interconnections with geographical neighbors provide access to new economic knowledge, thereby exerting knowledge spillover. In this regard, \cite{bahar2014} explained the knowledge spillover from neighbors by showing that the probability of a country developing a comparative advantage in an industry increases if a neighboring country has a comparative advantage in that same industry. Similarly, \cite{Boschma2016} suggested the neighbors sharing borders as the source of regional diversification at the sub-national level, demonstrating the specialization pattern of geographical neighbors shapes the local capability to develop new industries in a region. In a recent paper, \cite{Jun2020} pointed out that knowledge spillover from neighboring exporters is a significant predictor of increases in bilateral trade flows. In the similar context, \cite{Gao2021} showed that inter-regional spillover played a significant role in the industrial diversification of China's provinces.

In sum, the previous literature devoted academic efforts to analyze the inter-regional spillover from geographical neighbors as a function of physical distance and a direct link of the same industry itself. However, all geographical neighbors at the same distance is not homogeneous. For example, a port can make a different effect from an ordinary region that is not a port, since a port is a gateway so that products and industries nearby the port can be connected to the global market. Therefore, research that examines a inter-regional spillover with considering this heterogeneous feature of geographic neighbors. 

Meanwhile, the recent studies on multi-dimensional network spaces within a region have hinted at the empirical strategy of analyzing the spatial link between two different networks of related industries in a region and its neighbor. \cite{Jara-Figueroa2018} decomposed the related knowledge that workers bring into pioneer firms into two different dimensional spaces—the network of related industries and the network of related occupations. Then constructing the two relatedness indicators capturing industry and occupation knowledge, they estimated the spillover effect from different knowledge types on the survival rate of the firms within a region through logistic regression on the two relatedness indicators. In a recent paper, \cite{Catalan2020} formalized the concept of a bi-layered network to represent the interactions between scientific and technological knowledge and capabilities at the country level by suggesting a modified measurement called the scientific and technological cross-space. The point is that \cite{Jara-Figueroa2018} considered two exogenous types of knowledge and applied two relatedness indicators in a regression model of spillover separately, while \cite{Catalan2020} considered two endogenous types of knowledge and suggested a unified relatedness indicator of the interconnected space. 

\subsection{Port and regional development}
Undoubtedly, integrating transport and regional development is important in the literature of economic geography \citep{Fujita1999}. In particular, as maritime transport through ports has become the major transportation mode in international trade after the rise of GVC \citep{Amador2016,Jacobs2010}, recent studies on port and port regions have renewed the significance of the interdependence between transport and the economic performance of regions \citep{Bottasso2018,Ducruet2016,Moura2019,Zhao2020,Qi2020}. Starting from the idea of ports providing a comparative advantage to the economic activities of neighboring regions \citep{Fujita1996}, scholars tried to prove that ports function as an important infrastructure endowment to promote international trade and investment and facilitate regional development. \cite{Zhao2020}, for instances, argued the positive impact of port comprehensive strength on economic growth in local and neighboring cities in general but the regional differences in its spatial spillover affect either. \cite{Qi2020} probed the spatial spillover effects of logistics infrastructures including ports both at the national and regional levels in economic development, finding the positive dependence at the national level but the indecisive results in the regional level.

Most of these recent studies found evidence that the improvement in port infrastructural quality and logistics performance has in general a positive effect on regional economic growth, although the interdependence between port and regional activities may vary over time and space. In other words, existing studies have produced ambiguous results on spatial spillover from neighboring ports or only explained a broad correlation of regional growth with neighboring ports but still limited to understanding the underlying dynamics of mutual linkages. One main reason of this limitation is that most empirical studies investigate the influence of ports as aggregated measures of port activities or infrastructural characteristics, such as annual port traffic or throughput volume or the efficiency of port infrastructure (\citep{Ducruet2016}. 

A Port, however, is a window for a neighboring region that allows the interaction between global market and regional production. Since the list of products at a port is not just a piled products but a product, in which a knowledge for being connect to the global market is embedded. In this regard, a port is also a knowledge hub through which information and knowledge associated with the global market or the GPNs embodied in commodity flow \citep{Bottasso2018,Ducruet2016,Moura2019,Zhao2020,Qi2020}. With consideration of a port as a knowledge hub, \cite{Ducruet2016} investigated the links between commodity types of port traffic and economic activity types in port regions. Although their study succeeded in embracing the contribution of neighboring ports as sources of knowledge spillover, it didn't link the product structure of port and that of a neighboring region at product level, mainly because of data availability. 

Based on the literature review on inter-industry spillovers from related industries and on influences of nearby ports, there is the discrepancy between the practical importance and academic endeavors in looking at the two spillover channels in regional diversification by integrating ports as a source of knowledge. This study, therefore, explores the following questions to fill the gap in the previous literature: how do regions acquire the knowledge and capability that need to sustain and diversify their economic activities? Specifically, how do the international flows of commodities through neighboring ports contribute to regional diversification and resilience? Between spillover from a port (source from outside of a region) and other related industries within the region (source from inside of a region), which one is the prime knowledge source of industrial diversification of a sub-national region? Our expected contribution to the literature on economic geography and related diversification concerns the role of neighboring ports as a special case of neighbors exerting inter-regional spillover based on their own product spaces. Also, our research can nuisance to bridge the literature on the GPNs and that on the principle of relatedness regarding the economic development of sub-national regions.

\section{Methodology}

\subsection{Spatial unit of analysis: port and port region}
A port can be defined as a maritime facility that has wharves or loading areas, where ships load and discharge cargo. This study considers 40 export ports of Korea that are authorized by relevant regional offices, except for several tiny coastal ports, fishing ports, and the subject as a sum of all the new ports’ export value. With situating on a sea coast, its pier length is, for example, around 27,407m in the case of Incheon Port, which is one of the biggest ports in Korea. 

\begin{figure}[!htbp]
  \centering
  \includegraphics[width=0.45\textwidth]{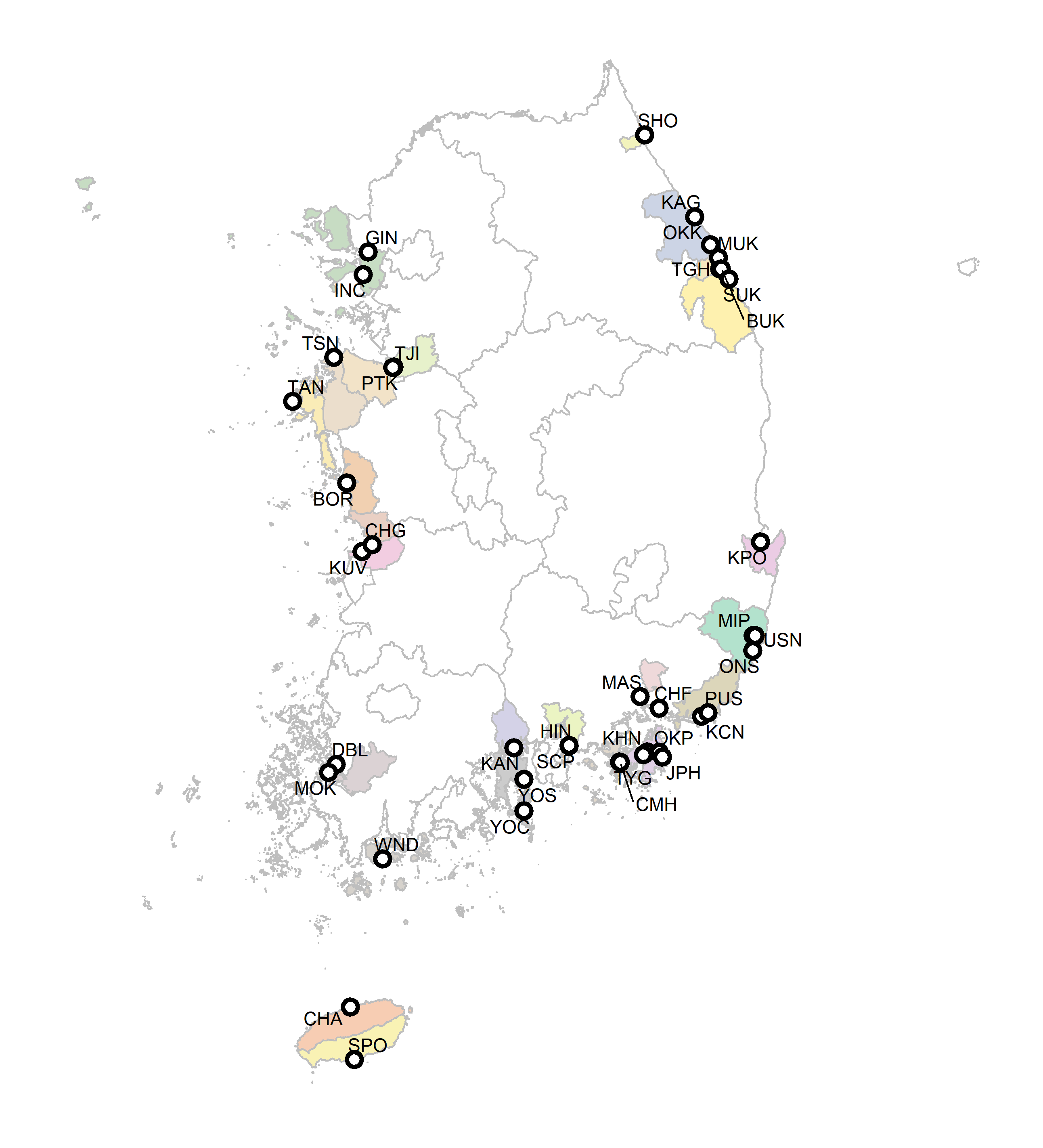}
  \caption{The scope of analysis, 40 ports and 26 port regions in Korea (see \textbf{Table~\ref{tab:the sample}} for further detail on the sample)}
  \label{fig:map}
\end{figure}

We define a port region as an \textit{si-gun-gu} level administrative district that has the port in their area. The average area of port regions in our analysis is around 625 $km^2$ and the export production in 26 port regions accounts for 56.4 \% of the total and deals with 1,224 of all 1,241 commodities. \textbf{Figure~\ref{fig:map}} shows the sample scope of analysis in this study. In the figure, black circles represent ports and shaded areas show the localization of port regions within the sample. 

From the 1960s, the Korean government strategically invested to develop ports associating with the export-oriented development plans. Considering that Korea is located at the center of Northeast Asian economies with its function as a logistics hub in the region, the role of ports has been crucial in the development strategies of Korea~\citep{jung2011economic}. Thanks to the growth potential with its location and the second wave of globalization, Korean ports have been growing over time and for example, around 13 million tons of cargo was transported through ports in Korea only in November 2021. As shown in \textbf{Figure~\ref{fig:map}}, moreover, the port regions in our analysis are geographically scattered over all three coasts in the Korean peninsula. Busan port, which locates in the Southern coast of Korean peninsula, has been regarded the representative gateway of Korean product towards the global market, but other ports such as Incheon port in the Western coast or Ulsan and Pohang port in Eastern coast are also handling more or similar massive quantity of cargo. In addition, according to \cite{lee2014bulkport}, the concentration of bulk cargos is decreasing over time in Korea, implying that the distribution of ports is not that skewed in Korea. 

There might be a potential concern that a region with the free trade zone shows distinctive feature and should be treated differently. However, according to \cite{park2008ftz}, upon the second wave of globalization, the benefits of free trade zone including tax benefit and low tariff decreased for firms that locate in the zone. Thus, considering that our data is made after the 2000s, we didn't distinguish the regions with the free trade zone from regions without them.

\subsection{Data}
We use two sets of export data for 2006 to 2020 in Korea: port data and regional data. Port data consists of the annual freight shipment value for each product at each port, including information on destination countries. Regional data are defined at the city or county level in Korea and consists of the annual production value of each product in each region shipped to each port. The data are extracted from Trade Statistics Service by Korea Customs Services (TRASS) Trade Statistics Service, which provides detailed trade statistics based on the Harmonized System (HS) four-digit aggregation (rev. 2017) for 1,241 commodities. For port data, the original sample of regional data consists of 160 cities and counties in Korea. Among them, we extract the sample of this study based on the location of ports: 26 regions out of the original sample are classified as port regions and matched with neighboring ports \citep{Bottasso2014, Ducruet2016}.

We also use the data on the Product Complexity Index (PCI) from MIT’s Observatory of Economic Complexity, measuring the knowledge intensity of products using world trade data \citep{Simoes2011, Hidalgo2009}. The PCI ranks not only the sophistication but also the diversity of the knowledge that is required to produce a product. For example, the PCI of a car is higher than that of a textile. The PCI data that we use distinguishes 1,060 products based on the HS four-digit classification (rev. 2002), so we use the conversion table from the UN Trade Statistics to convert and combine the HS 2002 codes into the HS 2017 codes of our sample.

\subsection{Product relatedness in production and transport}
How do neighboring ports affect the regional diversification of a port region?  Specifically, does the knowledge embodied in the commodity flows of neighboring ports influence the development of new industries in a port region? To explore our research question, the analytic framework of the \textit{Principle of Relatedness} is applied in this study \citep{hidalgo2018principle}. The \textit{Principle of Relatedness} is an outcome-based approach regarding the spatially concentrated knowledge and the likelihood that a region enters an economic activity related to a new product (i.e., the development of a new industry), based on the co-occurrence of economic activities related to ''similar products''. For its methodological means, first, product space is introduced, which is a network mapping the co-occurrence pattern of all products in a single space by calculating the proximity between products traded in an economy or a world. Second, product relatedness is brought in as a measure of capturing the knowledge and capability of a region to implement economic activities related to a specific product, depending on the number of related activities present in that location.

This study concerns two decomposed activities of export: producing commodities in regions and transporting them through ports. Therefore, when following the \textit{Principle of Relatedness} in this study, the most important starting point is to define what similar products are in each case. For regional data, similar products represent two products sharing production capabilities as both tend to be produced nearby in a region. For port data, similar products refer to the products that are likely to be transported at the same port based on similar infrastructure, institutions, and technology. Consequently, we should draw separate product spaces in production and transportation, computing the proximity for port and regional data following \cite{Hidalgo2007}. Here, the proximity between products in regional production $\phi$ or port activities $\Phi$ is a proxy of their similarity, measuring the shared knowledge and capabilities to produce or transport those products.

Mathematically, the proximity between products corresponds to the minimum of the pairwise conditional probability that a location exports both products with a revealed comparative advantage (RCA). Here, the $RCA$ of a region $r$ or port $p$ in a product $i$ measures whether $r$ or $p$ exports more of the product $i$ than the average, as a share of its total exports of Korea. By aggregating observations of the original data for 1,241 products from 2006 to 2020, we calculate the RCA following \cite{Balassa1965}:
\begin{equation}
RCA_{r, i} = \left. \frac{x_{r, i}}{\sum_{i}^{}{x_{r, i}}} \middle/ \frac{\sum_{r}^{}{x_{r, i}}}{\sum_{r, i}^{}{x_{r, i}}} \right.  \quad \textrm{and} \quad    
RCA_{p, i} = \left. \frac{X_{p, i}}{\sum_{i}^{}{X_{p, i}}} \middle/ \frac{\sum_{r}^{}{X_{p, i}}}{\sum_{p, i}^{}{X_{p, i}}} \right.
\label{eq:RCA}
\end{equation}
where $x_{r, i}$ and $X_{p, i}$ are matrices summarizing the export value (in US dollars) of a region $r$ and port $p$ in a product $i$. $RCA_i$ allows us to say that a region $r$ or port $p$ has the comparative advantage in a product $i$ (i.e., the industry $i$ is active) only when $RCA_i$ is greater or equal to 1.

\begin{figure}[!t]
\centering
\includegraphics[width=1\textwidth]{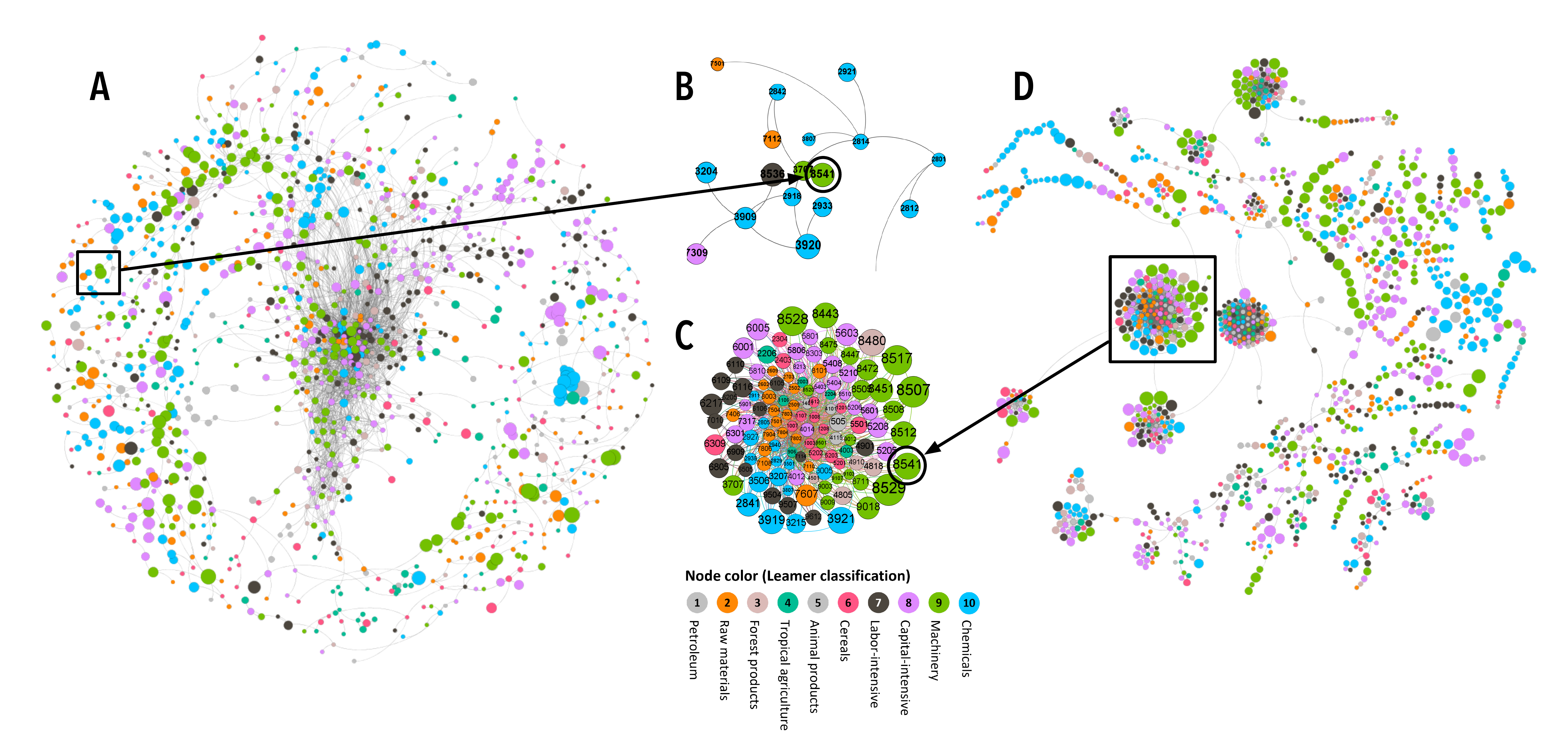}
\caption{Product space of Regions (A) and Ports (D) in period 2006-2020. B and C are ego networks of the product of four-digit HS code 8541, which is diodes, transistors, similar semiconductor devices. The node size represents the export value of the product in a relative scale, and the node color shows its classification as originally proposed by \cite{Leamer1984} and revised by \cite{Hidalgo2007}. By comparing the two different ego networks (B and C) of the same product (8541), we can indicate that the product space of region and of port show different structure.}
\label{fig:productspace_main}
\end{figure}

\begin{table}[!htbp]
  \centering
  \caption{Most similar products to the product of HS code 8541 (diodes, transistors, similar semiconductor devices) by production proximity (top) and by transport proximity (bottom)}
    \resizebox{\textwidth}{!}{\begin{tabular}{llll}
    \toprule
    \multicolumn{4}{c}{\textbf{Production proximity (in port regions)}} \\
    \midrule
    \multicolumn{1}{c}{\textbf{HS section}} & \multicolumn{1}{c}{\textbf{HS code}} & \textbf{Description} & \multicolumn{1}{c}{\textbf{Proximity}} \\
    \midrule
    \multicolumn{1}{c}{6} & \multicolumn{1}{c}{3707} & Chemical preparations for photographic uses & \multicolumn{1}{c}{0.72 } \\
    \multicolumn{1}{c}{16} & \multicolumn{1}{c}{8542} & Electronic integrated circuits & \multicolumn{1}{c}{0.69 } \\
    \multicolumn{1}{c}{18} & \multicolumn{1}{c}{9002} & Lenses, mirrors and other optical elements & \multicolumn{1}{c}{0.69 } \\
    \multicolumn{1}{c}{16} & \multicolumn{1}{c}{8534} & Circuits; printed & \multicolumn{1}{c}{0.68 } \\
    \multicolumn{1}{c}{13} & \multicolumn{1}{c}{7011} & Glass envelopes and glass parts  & \multicolumn{1}{c}{0.67 } \\
    \multicolumn{1}{c}{15} & \multicolumn{1}{c}{7413} & Copper; wire and the like & \multicolumn{1}{c}{0.67 } \\
    \multicolumn{1}{c}{16} & \multicolumn{1}{c}{8443} & Printing machinery & \multicolumn{1}{c}{0.67 } \\
    \multicolumn{1}{c}{16} & \multicolumn{1}{c}{8523} & Discs, tapes, smart cards & \multicolumn{1}{c}{0.66 } \\
    \multicolumn{1}{c}{6} & \multicolumn{1}{c}{2931} & Other organo-inorganic compounds & \multicolumn{1}{c}{0.66 } \\
    \multicolumn{1}{c}{16} & \multicolumn{1}{c}{8525} & Transmission apparatus for radio-broadcasting or television & \multicolumn{1}{c}{0.66 } \\
    \multicolumn{1}{c}{16} & \multicolumn{1}{c}{8532} & Electrical capacitors & \multicolumn{1}{c}{0.66 } \\
    \multicolumn{1}{c}{6} & \multicolumn{1}{c}{3208} & Paints, varnishes & \multicolumn{1}{c}{0.65 } \\
    \multicolumn{1}{c}{6} & \multicolumn{1}{c}{2810} & Oxides of boron & \multicolumn{1}{c}{0.64 } \\
    \multicolumn{1}{c}{18} & \multicolumn{1}{c}{9031} & Measuring or checking instruments and machines & \multicolumn{1}{c}{0.64 } \\
    \multicolumn{1}{c}{6} & \multicolumn{1}{c}{2942} & Organic compounds & \multicolumn{1}{c}{0.63 } \\
    \midrule
          &       &       &  \\
    \midrule
    \multicolumn{4}{c}{\textbf{Transport proximity (in port activiteis)}} \\
    \midrule
    \multicolumn{1}{c}{\textbf{HS section}} & \multicolumn{1}{c}{\textbf{HS code}} & \textbf{Description} & \multicolumn{1}{c}{\textbf{Proximity}} \\
    \midrule
    \multicolumn{1}{c}{1} & \multicolumn{1}{c}{505} & Skins and other parts of birds with feathers, down & \multicolumn{1}{c}{1.00 } \\
    \multicolumn{1}{c}{2} & \multicolumn{1}{c}{1003} & Barley & \multicolumn{1}{c}{1.00 } \\
    \multicolumn{1}{c}{2} & \multicolumn{1}{c}{1007} & Grain sorghum & \multicolumn{1}{c}{1.00 } \\
    \multicolumn{1}{c}{2} & \multicolumn{1}{c}{1008} & Buckwheat, millet and canary seeds; other cereals & \multicolumn{1}{c}{1.00 } \\
    \multicolumn{1}{c}{4} & \multicolumn{1}{c}{2204} & Wine of fresh grapes & \multicolumn{1}{c}{1.00 } \\
    \multicolumn{1}{c}{5} & \multicolumn{1}{c}{2602} & Manganese ores and concentrates & \multicolumn{1}{c}{1.00 } \\
    \multicolumn{1}{c}{6} & \multicolumn{1}{c}{2829} & Chlorates and perchlorates & \multicolumn{1}{c}{1.00 } \\
    \multicolumn{1}{c}{7} & \multicolumn{1}{c}{3921} & Plastic plates, sheets, film, foil & \multicolumn{1}{c}{1.00 } \\
    \multicolumn{1}{c}{8} & \multicolumn{1}{c}{4115} & Composition leather with a basis of leather  & \multicolumn{1}{c}{1.00 } \\
    \multicolumn{1}{c}{10} & \multicolumn{1}{c}{4805} & Uncoated paper and paperboard  & \multicolumn{1}{c}{1.00 } \\
    \multicolumn{1}{c}{11} & \multicolumn{1}{c}{6105} & Shirts; men's or boys', knitted or crocheted & \multicolumn{1}{c}{1.00 } \\
    \multicolumn{1}{c}{12} & \multicolumn{1}{c}{6505} & Hats and other headgear & \multicolumn{1}{c}{1.00 } \\
    \multicolumn{1}{c}{15} & \multicolumn{1}{c}{7607} & Aluminum foil & \multicolumn{1}{c}{1.00 } \\
    \multicolumn{1}{c}{16} & \multicolumn{1}{c}{8529} & Transmission apparatus &\multicolumn{1}{c}{1.00 } \\
    \multicolumn{1}{c}{17} & \multicolumn{1}{c}{8711} & Motorcycles and cycles & \multicolumn{1}{c}{1.00 } \\
    \midrule
    \multicolumn{4}{p{40em}}{\textit{$^1$ For the production proximity, top 15 pairs of the total are presented.}} \\
    \multicolumn{4}{p{40em}}{\textit{$^2$ For the transport proximity, 15 among 122 pairs with $\Phi=1$ are selected at random since there are too many pairs with $\Phi=1$.}} \\
    \end{tabular}}%
  \label{tab:most similar}%
\end{table}%

Using $RCA_i$ and $RCA_j$, we can create 1,241 x 1,241 matrices of the proximity between products $i$ and $j$ for a region $r$ ($\phi$, \textit{production proximity}) and a port $p$ ($\Phi$, \textit{transport proximity}), respectively:
\begin{equation}
\phi_{i, j} = \frac{\sum_{r}^{}{M_{r, i}M_{r, j}}}{max\{k_{r,i},k_{r,j}\}} \quad \textrm{and} \quad    
\Phi_{i, j} = \frac{\sum_{p}^{}{M_{p, i}M_{p, j}}}{max\{K_{p,i},K_{p,j}\}}
\label{eq:proximity}
\end{equation}
where $\phi_{i, j}$ and $\Phi_{i, j}\in[0,1]$; $M$ is a binary matrix of whether $r$ or $p$ has the comparative advantage in a product or not, and $k$ and $K$ are variables of the ubiquity of a product as the number of regions or ports having the comparative advantage in a product (i.e., $k_{r,i}=\sum_{}^{}{M_{r, i}}$ and $K_{p,j}=\sum_{}^{}{M_{p, i}}$ ). The proximity $\phi_{i, j}$ or $\Phi_{i, j}$ is one when products $i$ and $j$ always co-occur in the export data, while $\phi_{i, j}$ or $\Phi_{i, j}$ close to zero indicates no co-occurrence between products due to dissimilarity. 

\textbf{Table~\ref{tab:most similar}} shows an example of the most similar products to the product HS code 8541 (diodes, transistors, similar semiconductor devices) in terms of production proximity $\phi$ and transport proximity $\Phi$. As we expect, $\phi$ and $\Phi$ measure the different types of product similarity in production and transport. The top 15 products of production proximity around the product 8541 depict domestic value chains of producing semiconductor devices, including 6 products in the same family of industry and 9 products related to raw materials and machinery. Most similar products by transport proximity, on the other hand, vary from animal products to motorcycles. This result mainly comes from the containerization of semiconductor devices as general cargo. As shown in \textbf{Table~\ref{tab:cr}}, the containerization ratios of semiconductor devices and other proximate products have almost 1 in the case of Korean maritime transport. We can also see this distinct characteristics of transport proximity from \textbf{Figures \ref{fig:productspace_main}}. \textbf{Figures \ref{fig:productspace_main} A} depicts the product space of port regions, while \textbf{Figures \ref{fig:productspace_main} B} depicts that of ports. The example of product HS code 8541 tells us that the way of export including containerization gives effect on which port a product heads to export. Further information on two product spaces can be found in Appendix. 

Finally, we compute the product relatedness by year as the average proximity of a new potential product (i.e., a new industry) to the current structure of active export activities in a location \citep{Hidalgo2007, Hartmann2021}. Using two proximity matrices, we define two measures of product relatedness: the product relatedness of a product $i$ in a region $r$ in year $t$ $\omega_{r,i}^t$ and the product relatedness of a product $i$ in a port $r$ in year $t$ $\Omega_{p,i}^t$  as the following equation:
\begin{equation}
\omega_{r,i}^t = \frac{\sum_{j}^{}{M_{r, i}^{t}\phi_{i, j}}}{\sum_{j}^{}{\phi_{i, j}}} \quad \textrm{and} \quad    
\Omega_{p,i}^t = \frac{\sum_{j}^{}{M_{p, i}^{t}\Phi_{i, j}}}{\sum_{j}^{}{\Phi_{i, j}}}
\label{eq:density}
\end{equation}
where $M^t$ is 1 when $RCA^t$ is greater and equal to 1 in year $t$. The product relatedness $\omega_{r,i}^t$ or $\Omega_{p,i}^t$ is a value between 0 and 1, where $\omega_{r,i}^t$ or $\Omega_{p,i}^t$ close to one indicates that a region $r$ or port $p$ has abundant knowledge and capability of production or transport related to product $i$ in year $t$.

Back to our research question, the product relatedness in port activities $\Omega_{p,i}^t$ could be an explanatory variable to examine the spatial link between regional diversification and the knowledge embodied in the commodity flows of neighboring ports. For this purpose, we combine the port data and the selected regional data for port regions to create the port-region matched data based on \textbf{Table~\ref{tab:the sample}}. After matching, we rule out observations with missing values of $\Omega_{p,i}^t$ mainly generated because the port does not deal with some products manufactured in the port region\footnote{To compensate for the loss of information caused by ruling out such missing values, we consider the number of ports where each region ships each product as a control variable to the regression model.}. For the sample of 26 port regions, \textbf{Table~\ref{tab:statistics}} presents summary statistics of the original regional data and port-region matched data.

\subsection{Econometric model}
With the sample of 26 port regions, we conduct a multivariate probit regression to estimate whether product relatedness predicts increases in the probability of developing a new industry. We apply a two-way fixed-effect regression including year dummies and region dummies, with a standard error term allowing for within-cluster correlation of products to be robust under any product-specific characteristics\footnote{Although a three-way fixed-effect regression model, including product-fixed effect term as well, would be an good alternative in this study \citep{Jun2020, Catalan2020}, we choose a two-way fixed-effect model due to the limitation of the number of observations. Instead, we include the ubiquity k calculated from the product space and the product complexity index derived from the world trade data to control product characteristics of importance in this context and use a standard error term allowing for within-cluster correlation of products to be robust.}. For the model, we follow the tradition of relatedness literature  \citep{bahar2014,Boschma2016,Gao2021,Hausmann2007} and design the models for a two-step estimation by considering the spatial link between the ports’ and port regions’ product spaces. The first step is to estimate the effect of product relatedness in regional production ($\omega$) on developing a new industry, indicating knowledge spillover in own product space. In the second step, the model estimates the additional contribution of product relatedness in port activities ($\Omega$) to developing a new industry using the port-region matched data. This contribution represents the knowledge spillover that occurs when two product spaces with different types of knowledge are spatially connected (i.e., cross-space). If $\Omega$ shows positive dependence, we call this effect \textit{the cross-space spillover} in this study.

Formally, our model for the first step is given by Eq.\ref{eq:model1}:
\begin{equation}
S_{r,i}^{t+2} = \beta_0 + \beta_1 \omega_{r,i}^t +\beta_2 k_{r,i}^t + \beta_3 TRM_{r,i}^t + \beta_4 PCI_{i}^t + \mu_t +\mu_r + \varepsilon_{r,i}^t
\label{eq:model1}
\end{equation}
where the dependent variable $S_{r,i}^{t+2}$ is a binary variable of whether a region jumps to the new industry after two years, used to be no comparative advantage in that industry before. Here, to avoid noises of temporary jumps, we restrict jumps subject to the forward and backward conditions~\citep{bahar2014,Gao2021}: a jump needs to keep comparative advantages for two years more after the year $t+2$ (i.e., $M_{r,i}$ from $t+2$ to $t+4$ has to be 1), and a jump also satisfies $M_{r,i}=0$ for a further two years before the year $t$\footnote{For the period 2008-2016, we apply both the forward and backward conditions, but we only apply the forward condition for 2017-2018 and the backward condition for 2006-2007 due to the data limitation.}. In this model, the main explanatory variable $\omega$ indicates how much a region has knowledge of similar products around a product $i$ in year $t$, and $k_{r,i}^t$ represents the ubiquity of a product $i$ in year $t$. $TRM_{r,i}^t$ represents the total number of ports where each product in each region is shipped by year, and $PCI_{i}^t$ represents the product complexity index for a product $i$ in year $t$. $\mu_t$ and $\mu_r$ are fixed-effect terms to control omitted year-specific and region-specific variable bias, and $\varepsilon_{r,i}^t$ is the error term. 

In the second step, two main variables derived from the product space at ports are added to the previous model: the product relatedness and ubiquity in port activities, $\Omega_{p,i}^t$ and $K_{p,i}^t$. The final model of our interest is shown as below:
\begin{equation}
S_{r,p,i}^{t+2} = \beta_0 + \beta_1 \omega_{r,i}^t +\beta_2 \Omega_{p,i}^t +\beta_3 k_{r,i}^t + \beta_4 K_{p,i}^t +\beta_5 TRM_{r,i}^t +\beta_6 PCI_{i}^t + \mu_t +\mu_r +\varepsilon_{r,p,i}^t
\label{eq:model2}
\end{equation}
where $\beta_2$ is an estimator of cross-space spillover from neighboring ports to a port region, and  $\beta_1$ is an estimator of knowledge spillover within the regional product space.

\section{Results}
This section provides the regression results by estimating several specifications of Eqs.\ref{eq:model1} and \ref{eq:model2}. Based on our research questions, we divide the sample into groups and conduct group-wise comparisons of the estimates to verify factors to make differences in the degree of relevant spillover. Sections~\ref{ch:withinregion} and ~\ref{ch:crossspillover} present the empirical evidence of two spillover channels for regional diversification. Section~\ref{ch:resilience} discusses the role and significance of two spillover channels when a port region adapts to the change of trade landscape and external shocks.

\subsection{Knowledge spillover within the regional product space}
\label{ch:withinregion}
\textbf{Table \ref{tab:r}}  shows the baseline estimates of whether a region enters a new potential industry by Eq.\ref{eq:model1}, based on the local capability and knowledge of related industries in the current productive structure. Column (1) corresponds to the result considering all observations of the sample (see \textbf{Table \ref{tab:robust2}} for more regression estimates to test the robustness of our model). First, we find the positive dependency of $\omega$  on the development of a new industry, confirming the knowledge spillover within the regional product space~\citep{Hidalgo2007, Neffke2011, Gao2021}. This also supports the path dependency theory of regional diversification that a new industry emerges from existing industries in a region by recombining local capabilities related to them \citep{Boschma2013}. For other estimators, we discover that the ubiquity $k$ positively correlates with the development of a new industry while the product complexity $PCI$ shows a significant and negative correlation. This result suggests that the more ubiquitous and less complex product is a good candidate for a new potential industry in a port region. In this respect, we further investigate whether the knowledge spillover within the regional product space varies by the knowledge intensity of products and the industrial characteristics.

\FloatBarrier
\begin{table}[ht]
  \centering
  \caption{The development of a new potential industry by knowledge spillover within the regional product space}
    \resizebox{\textwidth}{!}{\begin{tabular}{lllllllll}
    \toprule
    \multicolumn{1}{c}{   } & \multicolumn{1}{c}{(1)} &       & \multicolumn{1}{c}{(2)} & \multicolumn{1}{c}{(3)} &       & \multicolumn{1}{c}{(4)} & \multicolumn{1}{c}{(5)} & \multicolumn{1}{c}{(6)} \\
\cmidrule{2-2}\cmidrule{4-5}\cmidrule{7-9}    \multicolumn{1}{c}{$S_{r,i}^{t+2}$} & \multicolumn{1}{c}{All} &       & \multicolumn{1}{c}{Low PCI} & \multicolumn{1}{c}{High PCI} &       & \multicolumn{1}{c}{Leamer~\newline{}1-6 } & \multicolumn{1}{c}{Leamer~\newline{} 7,8} & \multicolumn{1}{c}{Leamer~ \newline{}9,10} \\
\cmidrule{1-5}\cmidrule{7-9}    \multicolumn{1}{c}{$\omega_{r,i}^t$} & \multicolumn{1}{c}{7.041***} &       & \multicolumn{1}{c}{6.362***} & \multicolumn{1}{c}{8.137***} &       & \multicolumn{1}{c}{5.341***} & \multicolumn{1}{c}{7.426***} & \multicolumn{1}{c}{8.475***} \\
    \multicolumn{1}{c}{  } & \multicolumn{1}{c}{(0.342)} &       & \multicolumn{1}{c}{(0.417)} & \multicolumn{1}{c}{(0.535)} &       & \multicolumn{1}{c}{(0.521)} & \multicolumn{1}{c}{(0.698)} & \multicolumn{1}{c}{(0.568)} \\
    \multicolumn{1}{c}{$k_{r,i}^t$} & \multicolumn{1}{c}{0.007***} &       & \multicolumn{1}{c}{0.004**} & \multicolumn{1}{c}{0.010***} &       & \multicolumn{1}{c}{0.002} & \multicolumn{1}{c}{0.012***} & \multicolumn{1}{c}{0.007***} \\
    \multicolumn{1}{c}{  } & \multicolumn{1}{c}{(0.002)} &       & \multicolumn{1}{c}{(0.002)} & \multicolumn{1}{c}{(0.002)} &       & \multicolumn{1}{c}{(0.002)} & \multicolumn{1}{c}{(0.003)} & \multicolumn{1}{c}{(0.002)} \\
    \multicolumn{1}{c}{$PCI_{i}^t$} & \multicolumn{1}{c}{-0.058***} &       & \multicolumn{1}{c}{-0.080**} & \multicolumn{1}{c}{-0.032} &       & \multicolumn{1}{c}{-0.077**} & \multicolumn{1}{c}{0.025} & \multicolumn{1}{c}{-0.045} \\
    \multicolumn{1}{c}{  } & \multicolumn{1}{c}{(0.019)} &       & \multicolumn{1}{c}{(0.037)} & \multicolumn{1}{c}{(0.054)} &       & \multicolumn{1}{c}{(0.039)} & \multicolumn{1}{c}{(0.033)} & \multicolumn{1}{c}{(0.041)} \\
    \multicolumn{1}{c}{$TRM_{r,i}^t$ } & \multicolumn{1}{c}{0.156***} &       & \multicolumn{1}{c}{0.182***} & \multicolumn{1}{c}{0.138***} &       & \multicolumn{1}{c}{0.178**} & \multicolumn{1}{c}{0.196***} & \multicolumn{1}{c}{0.128***} \\
    \multicolumn{1}{c}{  } & \multicolumn{1}{c}{(0.021)} &       & \multicolumn{1}{c}{(0.041)} & \multicolumn{1}{c}{(0.023)} &       & \multicolumn{1}{c}{(0.085)} & \multicolumn{1}{c}{(0.028)} & \multicolumn{1}{c}{(0.027)} \\
    \multicolumn{1}{c}{$Constant$} & \multicolumn{1}{c}{-2.135***} &       & \multicolumn{1}{c}{-2.028***} & \multicolumn{1}{c}{-2.364***} &       & \multicolumn{1}{c}{-1.681***} & \multicolumn{1}{c}{-2.669***} & \multicolumn{1}{c}{-2.294***} \\
    \multicolumn{1}{c}{  } & \multicolumn{1}{c}{(0.105)} &       & \multicolumn{1}{c}{(0.136)} & \multicolumn{1}{c}{(0.177)} &       & \multicolumn{1}{c}{(0.181)} & \multicolumn{1}{c}{(0.189)} & \multicolumn{1}{c}{(0.202)} \\
    \multicolumn{1}{c}{Year dummies} & \multicolumn{1}{c}{yes} &       & \multicolumn{1}{c}{yes} & \multicolumn{1}{c}{yes} &       & \multicolumn{1}{c}{yes} & \multicolumn{1}{c}{yes} & \multicolumn{1}{c}{yes} \\
    \multicolumn{1}{c}{Region dummies} & \multicolumn{1}{c}{yes} &       & \multicolumn{1}{c}{ yes} & \multicolumn{1}{c}{ yes} &       & \multicolumn{1}{c}{ yes} & \multicolumn{1}{c}{ yes} & \multicolumn{1}{c}{ yes} \\
          &       &       &       &       &       &       &       &  \\
    \multicolumn{1}{c}{Observations} & \multicolumn{1}{c}{33,057 } &       & \multicolumn{1}{c}{15,246 } & \multicolumn{1}{c}{17,448 } &       & \multicolumn{1}{c}{7,879 } & \multicolumn{1}{c}{10,538 } & \multicolumn{1}{c}{14,501 } \\
    \multicolumn{1}{c}{Pseudo R$^2$} & \multicolumn{1}{c}{0.136} &       & \multicolumn{1}{c}{0.116} & \multicolumn{1}{c}{0.162} &       & \multicolumn{1}{c}{0.099} & \multicolumn{1}{c}{0.162} & \multicolumn{1}{c}{0.163} \\
    \multicolumn{1}{c}{log-likelihood} & \multicolumn{1}{c}{-14246} &       & \multicolumn{1}{c}{-7145} & \multicolumn{1}{c}{-6943} &       & \multicolumn{1}{c}{-4007} & \multicolumn{1}{c}{-4349} & \multicolumn{1}{c}{-5628} \\
    \multicolumn{1}{c}{Mean VIF} & \multicolumn{1}{c}{5.24} &       & \multicolumn{1}{c}{5.49} & \multicolumn{1}{c}{5.63} &       & \multicolumn{1}{c}{5.52} & \multicolumn{1}{c}{6.21} & \multicolumn{1}{c}{5.16} \\
    \bottomrule
    \multicolumn{9}{p{40em}}{\textit{$^1$ Standard errors are in parentheses.}} \\
    \multicolumn{9}{p{40em}}{\textit{$^2$ *** p$<$ 0.01, ** p$<$ 0.05, * p$<$ 0.1 }} \\
    \multicolumn{9}{p{43em}}{\textit{$^3$ Mean VIF represents the mean of uncentered variance inflation factors (VIFs) for all applied variables, detecting the collinearity.}} \\
    \end{tabular}}%
  \label{tab:r}%
\end{table}%

To analyze the effect of $\omega$ by levels of product complexity, the sample is divided into two groups based on the $PCI$ value: products below the sample mean of PCI are classified in the low PCI group, and otherwise in the high PCI group. Columns (2) and (3) report the regression estimates for each group. The coefficients of omega between groups suggest statistical differences by the chi-square test (the chi-square with one degree of freedom = 14.16, p = 0.0002): the effect of relatedness of high PCI products is larger than that of low PCI products on the development of a new industry. This result proves that the higher the knowledge intensity of the product, the more is it influenced by knowledge spillover from related products when a region develops a new industry~\citep{Rauch1999, Rauch2001, Jun2020}.  

In addition to product complexity, the degree of knowledge spillover could be determined by the industrial characteristics such as factors of production and technological sophistication. Although a measure of product relatedness mainly captures product-specific factors and is partly industry-specific \citep{Hausmann2007}, the development of a new industry, our interest, is often conditioned by such industrial characteristics. Therefore, we also aim to identify the broad pattern of the knowledge spillover by sophistication levels of industrial production \citep{Felipe2014}. For this reason, we classify products into 10 categories of the Leamer classification by relative factor intensities of capital, labor, and skills required for each category, following \cite{Hidalgo2007}\footnote{According to \cite{Hidalgo2007}, we can apply the revised version of the classification introduced by \cite{Leamer1984}: petroleum (Leamer 1), raw materials (Leamer 2), forest products (Leamer 3), tropical agriculture (Leamer 4), animal agriculture (Leamer 5), cereals (Leamer 6), labor-intensity (Leamer 7), capital-intensive (Leamer8), machinery (Leamer 9), chemicals (Leamer 10).}. Considering the characteristics of 10 categories, then, we split the sample into three groups by industry sophistication levels: the low group for primary industries and raw material sectors, the medium group for labor- and capital-intensive industries, and the high group for machinery and chemical sectors required high skills for production (i.e., knowledge-based industries).

In \textbf{Table \ref{tab:r}}, Columns (4) to (6) show the estimations by industry sophistication levels. We find that the coefficients of $\omega$ increases with the level of industry sophistication, and the differences between groups are statistically significant at a 10\% confidence level. This result confirms when a region is specialized more in sophisticated industries (e.g., machinery and chemicals), the current productive structure exerts a stronger spatial spillover effect to develop a new related industry. Together with the estimates by product complexity levels, our results extend the literature on regional diversification on relatedness by probing that the product relatedness of more complex products and sophisticated industries has a greater influence on the development of new industries in a region \citep{Jun2020}.

\subsection{Cross-space spillover from neighboring ports}
\label{ch:crossspillover}
In the previous section, our results explain more about the generality of regional diversification rather than the specialty from a port region point of view. Here, we use the sample of matching port regions with their neighboring ports and apply Eq. \ref{eq:model2}. \textbf{Table \ref{tab:rp}} reports the regression estimates of our main interest, the contribution of neighboring ports to developing a new industry in a port region. The first column of \textbf{Table \ref{tab:rp}} shows the result considering all observations of the matched sample (see also \textbf{Table \ref{tab:robust2}} for more regression results to check the robustness of our model). The effect of $\omega$ as the local capability to produce related products is still significant in developing a new industry, and the effect of $\Omega$ from neighboring ports is non-negligible either. Even under the product relatedness within its own product space as the main spillover channel, this result confirms the knowledge embodied in transporting goods at ports flows into its neighboring region and contributes to the diversification of regional productive structures. This result demonstrates the cross-space spillover: the commodity flows of ports could function as a source of the knowledge spillover for their port regions.

\FloatBarrier
\begin{table}[ht]
  \centering
  \caption{The development of a new potential industry by cross-space spillover from neighboring ports }
    \resizebox{\textwidth}{!}{\begin{tabular}{ccccccccc}
    \toprule
          & \multicolumn{1}{c}{(1)} &       & \multicolumn{1}{c}{(2)} & \multicolumn{1}{c}{(3)} &       & \multicolumn{1}{c}{(4)} & \multicolumn{1}{c}{(5)} & \multicolumn{1}{c}{(6)} \\
\cmidrule{2-2}\cmidrule{4-5}\cmidrule{7-9}   $S_{r,p,i}^{t+2}$ & \multicolumn{1}{c}{All} &       & \multicolumn{1}{c}{Low PCI} & \multicolumn{1}{c}{High PCI} &       & \multicolumn{1}{c}{Leamer 1-6 } & \multicolumn{1}{c}{Leamer 7,8} & \multicolumn{1}{c}{Leamer 9,10} \\
    \midrule
    $\omega_{r,i}^t$  & \multicolumn{1}{c}{7.833***} &       & \multicolumn{1}{c}{7.258***} & \multicolumn{1}{c}{8.634***} &       & \multicolumn{1}{c}{6.318***} & \multicolumn{1}{c}{8.258***} & \multicolumn{1}{c}{8.861***} \\
          & \multicolumn{1}{c}{(0.424)} &       & \multicolumn{1}{c}{(0.541)} & \multicolumn{1}{c}{(0.615)} &       & \multicolumn{1}{c}{(0.701)} & \multicolumn{1}{c}{(0.853)} & \multicolumn{1}{c}{(0.675)} \\
    $\Omega_{p,i}^t$  & \multicolumn{1}{c}{0.707***} &       & \multicolumn{1}{c}{0.701***} & \multicolumn{1}{c}{0.693***} &       & \multicolumn{1}{c}{0.769*} & \multicolumn{1}{c}{1.098***} & \multicolumn{1}{c}{0.557**} \\
          & \multicolumn{1}{c}{(0.175)} &       & \multicolumn{1}{c}{(0.224)} & \multicolumn{1}{c}{(0.243)} &       & \multicolumn{1}{c}{(0.466)} & \multicolumn{1}{c}{(0.354)} & \multicolumn{1}{c}{(0.221)} \\
    $k_{r,i}^t$      & \multicolumn{1}{c}{0.056***} &       & \multicolumn{1}{c}{0.037**} & \multicolumn{1}{c}{0.076***} &       & \multicolumn{1}{c}{0.061***} & \multicolumn{1}{c}{0.069***} & \multicolumn{1}{c}{0.059***} \\
    \multicolumn{1}{l}{} & \multicolumn{1}{c}{(0.012)} &       & \multicolumn{1}{c}{(0.015)} & \multicolumn{1}{c}{(0.017)} &       & \multicolumn{1}{c}{(0.023)} & \multicolumn{1}{c}{(0.018)} & \multicolumn{1}{c}{(0.018)} \\
    $K_{p,i}^t$     & \multicolumn{1}{c}{0.006***} &       & \multicolumn{1}{c}{0.004} & \multicolumn{1}{c}{0.009***} &       & \multicolumn{1}{c}{0.000 } & \multicolumn{1}{c}{0.010**} & \multicolumn{1}{c}{0.006*} \\
          & \multicolumn{1}{c}{(0.002)} &       & \multicolumn{1}{c}{(0.003)} & \multicolumn{1}{c}{(0.003)} &       & \multicolumn{1}{c}{(0.004)} & \multicolumn{1}{c}{(0.004)} & \multicolumn{1}{c}{(0.003)} \\
   $PCI_{,i}^t$  & \multicolumn{1}{c}{-0.010 } &       & \multicolumn{1}{c}{-0.018 } & \multicolumn{1}{c}{0.115 } &       & \multicolumn{1}{c}{-0.040 } & \multicolumn{1}{c}{0.026 } & \multicolumn{1}{c}{0.045 } \\
          & \multicolumn{1}{c}{(0.025)} &       & \multicolumn{1}{c}{(0.048)} & \multicolumn{1}{c}{(0.073)} &       & \multicolumn{1}{c}{(0.053)} & \multicolumn{1}{c}{(0.043)} & \multicolumn{1}{c}{(0.055)} \\
    $TRM_{r,i}^t$   & \multicolumn{1}{c}{0.111***} &       & \multicolumn{1}{c}{0.116***} & \multicolumn{1}{c}{0.112***} &       & \multicolumn{1}{c}{0.104} & \multicolumn{1}{c}{0.162***} & \multicolumn{1}{c}{0.086***} \\
          & \multicolumn{1}{c}{(0.022)} &       & \multicolumn{1}{c}{(0.040)} & \multicolumn{1}{c}{(0.026)} &       & \multicolumn{1}{c}{(0.086)} & \multicolumn{1}{c}{(0.030)} & \multicolumn{1}{c}{(0.029)} \\
    $Constant$ & \multicolumn{1}{c}{-3.001***} &       & \multicolumn{1}{c}{-2.893***} & \multicolumn{1}{c}{-8.001***} &       & \multicolumn{1}{c}{-6.168***} & \multicolumn{1}{c}{-8.272***} & \multicolumn{1}{c}{-2.718***} \\
          & \multicolumn{1}{c}{(0.249)} &       & \multicolumn{1}{c}{(0.387)} & \multicolumn{1}{c}{(0.538)} &       & \multicolumn{1}{c}{(0.706)} & \multicolumn{1}{c}{(0.734)} & \multicolumn{1}{c}{(0.332)} \\
    Year dummies & \multicolumn{1}{c}{yes} &       & \multicolumn{1}{c}{yes} & \multicolumn{1}{c}{yes} &       & \multicolumn{1}{c}{yes} & \multicolumn{1}{c}{yes} & \multicolumn{1}{c}{yes} \\
    Region dummies & \multicolumn{1}{c}{yes} &       & \multicolumn{1}{c}{ yes} & \multicolumn{1}{c}{ yes} &       & \multicolumn{1}{c}{ yes} & \multicolumn{1}{c}{ yes} & \multicolumn{1}{c}{ yes} \\
    \multicolumn{1}{r}{} &       &       &       &       &       &       &       &  \\
    Observations & \multicolumn{1}{c}{13,735 } &       & \multicolumn{1}{c}{6,173 } & \multicolumn{1}{c}{7,365 } &       & \multicolumn{1}{c}{2,852 } & \multicolumn{1}{c}{4,361 } & \multicolumn{1}{c}{6,435 } \\
    Pseudo R$^2$ & \multicolumn{1}{c}{0.206} &       & \multicolumn{1}{c}{0.185} & \multicolumn{1}{c}{0.229} &       & \multicolumn{1}{c}{0.162} & \multicolumn{1}{c}{0.206} & \multicolumn{1}{c}{0.241} \\
    log-likelihood & \multicolumn{1}{c}{-6138 } &       & \multicolumn{1}{c}{-2994 } & \multicolumn{1}{c}{-3064 } &       & \multicolumn{1}{c}{-1505 } & \multicolumn{1}{c}{-1963 } & \multicolumn{1}{c}{-2563 } \\
    Mean VIF & \multicolumn{1}{c}{8.33 } &       & \multicolumn{1}{c}{9.10 } & \multicolumn{1}{c}{4.85 } &       & \multicolumn{1}{c}{6.43 } & \multicolumn{1}{c}{5.74 } & \multicolumn{1}{c}{7.66 } \\
    \midrule
    \multicolumn{9}{p{40em}}{\textit{$^1$ Standard errors are in parentheses.}} \\
    \multicolumn{9}{p{40em}}{\textit{$^2$ *** p$<$ 0.01, ** p$<$ 0.05, * p$<$ 0.1 }} \\
    \multicolumn{9}{p{43em}}{\textit{$^3$ Mean VIF represents the mean of uncentered variance inflation factors (VIFs) for all applied variables, detecting the collinearity.}} \\
    \end{tabular}}%
  \label{tab:rp}%
\end{table}%

Next, we analyze the dependency of cross-space spillover on the level of product complexity and industry sophistication. Columns (2) and (3) of \textbf{Table \ref{tab:rp}} report the regression results for the low and high PCI groups. Interestingly, the estimators of $\Omega$, indicating cross-space spillover effects, are significantly positive in both groups but with no substantial differences between groups. This result demonstrates that the knowledge captured by the product relatedness in port activities only has a spillover effect in a broad sense, regardless of products’ complexities. This is quite different from the effect of $\omega$, the influence of knowledge spillover from regional capability in related industries. We expect this limited dependency is mainly because the knowledge required to produce highly complex products qualitatively differentiates from the knowledge to transport them. Specifically, cross-space spillover plays an extra role in developing a new industry only by providing information about whether the product is easy to export with other commodities to the global market (i.e., a port as a global hub of production sharing). As presented in Columns (4) to (6), we also yield similar results and implications to the result in Table 2, by comparing the three groups with different sophistication levels of industrial production.

The ubiquity in port activities $K$, indicating the number of ports with $RCA$ above 1 in exporting the product, also shows positive significance in the cases of high PCI products and the industries with medium and high sophistication levels. This result suggests that plenty of active ports connecting global and local markets for the potential product increase the probability of developing the new industry especially for complex and knowledge-based products, like electronic equipment and pharmaceutical materials. Meanwhile, the ubiquity in regional production $k$, indicating the number of port regions with $RCA$ above 1 in producing the export, shows the positive correlation with developing a new industry in all groups. The universal positive correlation indicates that more ubiquitous products in port regions are always good candidates for a new potential industry as other successful domestic exporters are considered as the benchmark for the industry.

\subsection{Product relatedness as an economic resilience}
\label{ch:resilience}
Concerning ports as gateways of international trade, our next question arises: When does the role of the knowledge embodied in exporting products through neighboring ports stand out, especially in terms of the regional diversification induced by the changes in the trade environment? \textbf{Table \ref{tab:rpp}} shows the results divided into three periods: the period of economic crisis (2007–2009), the period of recovery (2010–2013), and the period of post-crisis (2014–2018). Before getting to the main point, we point out that the impact of $\omega$ is remarkable during the recovery period. This implies that the local production capability to related industries could serve as not just the spillover channel but more the regional resilience to adapt to the new circumstances following the economic crisis. Then, we discover that the effects of $\Omega$, indicating cross-space spillover from neighboring ports, are significantly positive only in times after the global economic crisis and have a marginal difference in degrees between periods. This result suggests that the product relatedness in port activities complements the economic resilience of a port region after the economic shock, as neighboring ports have the cross-space spillover effect to inform about potential co-exported goods as the candidate industry.

\FloatBarrier
\begin{table}[ht]
  \centering
  \caption{The development of a new industry in times of economic crisis (2007-2009), recovery (2010-2013) and post-crisis (2014-2018)}
    \resizebox{\textwidth}{!}{\begin{tabular}{c m{10em} m{10em} m{10em}}
    \toprule
          & \multicolumn{1}{c}{(1)} & \multicolumn{1}{c}{(2)} & \multicolumn{1}{c}{(3)} \\
\cmidrule{2-4}    $S_{r,p,i}^{t+2}$ & \multicolumn{1}{c}{\makecell{Economic risis\\(2007-2009)}} & \multicolumn{1}{c}{\makecell{Recovery\\(2010-2013)}} & \multicolumn{1}{c}{\makecell{Post-crisis\\(2014-2018)}  } \\
    \midrule
    $\omega_{r,i}^t$ & \multicolumn{1}{c}{7.467***} & \multicolumn{1}{c}{9.544***} & \multicolumn{1}{c}{7.714***} \\
          & \multicolumn{1}{c}{(1.176)} & \multicolumn{1}{c}{(0.728)} & \multicolumn{1}{c}{(0.584)} \\
    $\Omega_{p,i}^t$ & \multicolumn{1}{c}{0.868} & \multicolumn{1}{c}{0.981**} & \multicolumn{1}{c}{0.563***} \\
    \multicolumn{1}{l}{} & \multicolumn{1}{c}{(0.600)} & \multicolumn{1}{c}{(0.440)} & \multicolumn{1}{c}{(0.177)} \\
    $k_{r,i}^t$     & \multicolumn{1}{c}{0.010**} & \multicolumn{1}{c}{0.004} & \multicolumn{1}{c}{0.004} \\
    \multicolumn{1}{l}{} & \multicolumn{1}{c}{(0.004)} & \multicolumn{1}{c}{(0.003)} & \multicolumn{1}{c}{(0.003)} \\
    $K_{r,i}^t$     & \multicolumn{1}{c}{0.019} & \multicolumn{1}{c}{0.048***} & \multicolumn{1}{c}{0.078***} \\
          & \multicolumn{1}{c}{(0.021)} & \multicolumn{1}{c}{(0.018)} & \multicolumn{1}{c}{(0.016)} \\
    $PCI_{i}^t$  & \multicolumn{1}{c}{0.038 } & \multicolumn{1}{c}{0.006 } & \multicolumn{1}{c}{-0.036 } \\
          & \multicolumn{1}{c}{(0.046)} & \multicolumn{1}{c}{(0.033)} & \multicolumn{1}{c}{(0.033)} \\
    $TRM_{r,i}^t$   & \multicolumn{1}{c}{0.133***} & \multicolumn{1}{c}{0.169***} & \multicolumn{1}{c}{0.071**} \\
          & \multicolumn{1}{c}{(0.037)} & \multicolumn{1}{c}{(0.026)} & \multicolumn{1}{c}{(0.030)} \\
    $Constant$ & \multicolumn{1}{c}{-7.027***} & \multicolumn{1}{c}{-8.453***} & \multicolumn{1}{c}{-2.985***} \\
          & \multicolumn{1}{c}{(1.008)} & \multicolumn{1}{c}{(0.689)} & \multicolumn{1}{c}{(0.301)} \\
    Year dummies & \multicolumn{1}{c}{yes} & \multicolumn{1}{c}{yes} & \multicolumn{1}{c}{yes} \\
    Region dummies & \multicolumn{1}{c}{ yes} & \multicolumn{1}{c}{ yes} & \multicolumn{1}{c}{ yes} \\
    \multicolumn{1}{r}{} &       &       &  \\
    Observations & \multicolumn{1}{c}{2,820 } & \multicolumn{1}{c}{4,748 } & \multicolumn{1}{c}{5,450 } \\
    Pseudo R$^2$ & \multicolumn{1}{c}{0.188} & \multicolumn{1}{c}{0.203} & \multicolumn{1}{c}{0.213} \\
    log-likelihood & \multicolumn{1}{c}{-1269.959} & \multicolumn{1}{c}{-2115.669} & \multicolumn{1}{c}{-2543.948} \\
    Mean VIF & \multicolumn{1}{c}{7.10 } & \multicolumn{1}{c}{5.20 } & \multicolumn{1}{c}{9.48 } \\
    \midrule
    \multicolumn{4}{p{40.26em}}{\textit{$^1$Standard errors are in parentheses}} \\
    \multicolumn{4}{p{40.26em}}{\textit{$^2$*** p$<$ 0.01, ** p$<$ 0.05, * p$<$ 0.1 }} \\
    \multicolumn{4}{p{40.26em}}{\textit{$^3$ Mean VIF represents the mean of uncentered variance inflation factors (VIFs) for all applied variables, detecting the collinearity.}} \\
    \end{tabular}}%
  \label{tab:rpp}%
\end{table}%

By comparing the results of Columns (2) and (3), we find a statistically significant difference in the coefficients of $TRM$, measuring how many ports each product in each region is shipped to, between the recovery and post-crisis periods:  the increase of $TRM$ during the post-crisis period (2014--2018) is less effective than that during the years right after the crisis. This result has interesting implications in that, in general, port regions are linked with multiple ports for shipping each product when diversifying export destinations to multiple continents. In other words, when developing a new industry, a port region has less incentive to target products with various global destinations than before the crisis, proving Korea's export industry is also being affected by a slowdown in globalization and a loosened GVC \citep{Bailey2014,Lund2020} (see \textbf{Table \ref{tab:postcrisis}}, applying the variable $DES$ of representing the total number of destination continents of a product $i$ at a port $p$, instead of $TRM$, and confirming the significantly negative impact of $DES$ on developing a new industry only during the post-crisis).

\newpage
\section{Conclusion}
The global economy is thrown into chaos due to multiple events occurring simultaneously in the current time after the 2008 financial crisis, and all countries and regions have been experiencing unprecedented challenges from such a prolonged crisis. However, the world is not flat. Some regions show discontinuity in their socio-economic features shaping economic resilience – how much the regional economy sustains its normality despite the exogenous shock. Having a port nearby, for instance, is the discontinuous feature considering that ports are not only physical gateways linking the global market with local production but also knowledge hubs through which information and knowledge embodied in commodity flow through international production networks. However, little attention has been paid to embracing the distinctive characteristics of ports and ports' activities into empirical researches in economic complexity and economic geography. Scholars have rather studied the inter-regional spillover effects on the industrial diversification and economic development from their neighbors as homogeneous geographic units by their physical distance.


We studied the effects of inter-regional spillover of neighboring ports on the industrial diversification of regions. First, we found that the product space of ports shows their distinct characteristics mainly because of how each product deals with in logistics system. This implies that similar products or industries on the logistics side can differ from those on the production side in terms of economic complexity. In other words, there may exist different types of capabilities behind the product space of ports and regions. Second, our econometric estimates show that regions are more likely to enter a new industry when they already have the related industries in their local productive structure, confirming \cite{Hidalgo2007, Neffke2011, Boschma2013, Gao2021}. Interestingly, we also proved that regions' entering a new industry can be catalyzed by neighboring ports having competitiveness in the related products to the new industry. In addition, by splitting our sample over PCI and leamer's classification, we found that the effect of inter-industry spillover within the same regions increases with product complexity and technological sophistication of the new potential industry, while the effect of cross-space inter-regional spillover is the strongest for the new potential industry with medium-level sophistication(i.e., labor- and capital-intensive industries). This result also supports the existence of different micro-channel in spillovers between ports and regions. Finally, we confirmed that this cross-space inter-regional spillover counts for regional economies in developing new industries during the recovery period after the economic crisis, suggesting port activities positively contribute to regional resilience but weakened their influences recently due to a loosened global value chain.

However, our research is limited in figuring out the micro-mechanisms of the two different types of spillover. Behind the inter-industry spillover within the region, there might be the labor flow from the related industries to a new industry~\citep{Jara-Figueroa2018}, knowledge flows among product lines, and further social capability~\citep{abramovitz1986catching}, technological capability~\citep{kim1999building}, or institutions~\citep{acemoglu2005institutions} that support the already existing related industries. On the other hand, behind the scene of cross-space inter-regional spillovers from neighboring ports to a region, there might be knowledge embedded commodity flows, reprocessing of imported products, or trade information flows among traders. 

Despite our limitation, our results shed light on the cross-space spillover among different geographical dimensions and suggest the role of ports in knowledge spillover. This research tells us that the various types of knowledge spillover channels are engaged in regional industrial diversification and each channel plays a role with different intensity over product complexity and technological sophistication, suggesting that more delicate and targeted policy is required for regional industrial diversification.

\section*{Acknowledgement}
This project is funded by the National Research Foundation of Korea (NRF-2022R1A2C1012895). We also acknowledge the support of the Inha University.

\theendnotes

\linespread{1.5}
\newpage
\bibliographystyle{dcu}
\biboptions{authoryear}
\bibliography{reference.bib}

\newpage
\linespread{1.0}
\appendix
\section*{Appendix}

\setcounter{table}{0}
\setcounter{figure}{0} 
\renewcommand{\thetable}{A\arabic{table}}
\renewcommand{\thefigure}{A\arabic{figure}}

\begin{table}[htb!]
  \centering
  {\small
  \caption{The sample information}
    \begin{tabular*}{\textwidth}{@{\extracolsep{\fill}}lllll}
    \toprule
          & \multicolumn{2}{c}{\textbf{Port}} & \multicolumn{2}{c}{\textbf{Port region}} \\
\cmidrule{2-5}    \multicolumn{1}{c}{\textbf{Coast}} & \multicolumn{1}{c}{\textbf{Code}} & \multicolumn{1}{c}{\textbf{Name}} & \multicolumn{1}{c}{\textbf{Code}} & \multicolumn{1}{c}{\textbf{Name}} \\
    \midrule
    \multicolumn{1}{c}{south} & \multicolumn{1}{c}{KCN} & \multicolumn{1}{c}{Ghmcheon} & \multicolumn{1}{c}{RBUS} & \multicolumn{1}{c}{Busan} \\
    \multicolumn{1}{c}{south} & \multicolumn{1}{c}{PSN} & \multicolumn{1}{c}{Busan new} & \multicolumn{1}{c}{RBUS} & \multicolumn{1}{c}{Busan} \\
    \multicolumn{1}{c}{south} & \multicolumn{1}{c}{PUS} & \multicolumn{1}{c}{Busan} & \multicolumn{1}{c}{RBUS} & \multicolumn{1}{c}{Busan} \\
    \multicolumn{1}{c}{south} & \multicolumn{1}{c}{MAS} & \multicolumn{1}{c}{Masan} & \multicolumn{1}{c}{RCHW} & \multicolumn{1}{c}{Changwon} \\
    \multicolumn{1}{c}{south} & \multicolumn{1}{c}{CHF} & \multicolumn{1}{c}{Jinhae} & \multicolumn{1}{c}{RCHW} & \multicolumn{1}{c}{Changwoni} \\
    \multicolumn{1}{c}{south} & \multicolumn{1}{c}{KHN} & \multicolumn{1}{c}{Kohyeon} & \multicolumn{1}{c}{RKJE} & \multicolumn{1}{c}{Geoje} \\
    \multicolumn{1}{c}{south} & \multicolumn{1}{c}{OKP} & \multicolumn{1}{c}{Okpo} & \multicolumn{1}{c}{RKJE} & \multicolumn{1}{c}{Geoje} \\
    \multicolumn{1}{c}{south} & \multicolumn{1}{c}{JPH} & \multicolumn{1}{c}{Jangseungpo} & \multicolumn{1}{c}{RKJE} & \multicolumn{1}{c}{Geoje} \\
    \multicolumn{1}{c}{south} & \multicolumn{1}{c}{KJE} & \multicolumn{1}{c}{Geoje} & \multicolumn{1}{c}{RKJE} & \multicolumn{1}{c}{Geoje} \\
    \multicolumn{1}{c}{south} & \multicolumn{1}{c}{TYG} & \multicolumn{1}{c}{Tongyeong} & \multicolumn{1}{c}{RTYG} & \multicolumn{1}{c}{Tongyeong} \\
    \multicolumn{1}{c}{south} & \multicolumn{1}{c}{CMH} & \multicolumn{1}{c}{Chungmu} & \multicolumn{1}{c}{RTYG} & \multicolumn{1}{c}{Tongyeong} \\
    \multicolumn{1}{c}{south} & \multicolumn{1}{c}{SCP} & \multicolumn{1}{c}{Samcheonpo} & \multicolumn{1}{c}{RSCN} & \multicolumn{1}{c}{Sacheon} \\
    \multicolumn{1}{c}{south} & \multicolumn{1}{c}{HIN} & \multicolumn{1}{c}{Jinju} & \multicolumn{1}{c}{RSCN} & \multicolumn{1}{c}{Sacheon} \\
    \multicolumn{1}{c}{south} & \multicolumn{1}{c}{DBL} & \multicolumn{1}{c}{Daebul} & \multicolumn{1}{c}{RYAM} & \multicolumn{1}{c}{Yeongam} \\
    \multicolumn{1}{c}{south} & \multicolumn{1}{c}{MOK} & \multicolumn{1}{c}{Mokpo} & \multicolumn{1}{c}{RMOK} & \multicolumn{1}{c}{Mokpo} \\
    \multicolumn{1}{c}{south} & \multicolumn{1}{c}{YOS} & \multicolumn{1}{c}{Yeosu} & \multicolumn{1}{c}{RYOS} & \multicolumn{1}{c}{Yeosu} \\
    \multicolumn{1}{c}{south} & \multicolumn{1}{c}{WND} & \multicolumn{1}{c}{Wando} & \multicolumn{1}{c}{RWND} & \multicolumn{1}{c}{Wando$^1$} \\
    \multicolumn{1}{c}{south} & \multicolumn{1}{c}{KAN} & \multicolumn{1}{c}{Gwangyang} & \multicolumn{1}{c}{RGWY} & \multicolumn{1}{c}{Gwangyang} \\
    \multicolumn{1}{c}{south} & \multicolumn{1}{c}{YOC} & \multicolumn{1}{c}{Yeocheon} & \multicolumn{1}{c}{RGWY} & \multicolumn{1}{c}{Gwangyang} \\
    \multicolumn{1}{c}{south} & \multicolumn{1}{c}{SPO} & \multicolumn{1}{c}{Seogwipo} & \multicolumn{1}{c}{RSPO} & \multicolumn{1}{c}{Seogwipo} \\
    \multicolumn{1}{c}{south} & \multicolumn{1}{c}{CHA} & \multicolumn{1}{c}{Jeju} & \multicolumn{1}{c}{RJEJ} & \multicolumn{1}{c}{Jeju} \\
    \multicolumn{1}{c}{east} & \multicolumn{1}{c}{KAG} & \multicolumn{1}{c}{Gangneung} & \multicolumn{1}{c}{RKAG} & \multicolumn{1}{c}{Gangneung} \\
    \multicolumn{1}{c}{east} & \multicolumn{1}{c}{TGH} & \multicolumn{1}{c}{Donghae} & \multicolumn{1}{c}{RTGH} & \multicolumn{1}{c}{Donghae} \\
    \multicolumn{1}{c}{east} & \multicolumn{1}{c}{MUK} & \multicolumn{1}{c}{Mukho} & \multicolumn{1}{c}{RTGH} & \multicolumn{1}{c}{Donghae} \\
    \multicolumn{1}{c}{east} & \multicolumn{1}{c}{BUK} & \multicolumn{1}{c}{Bukpyeong} & \multicolumn{1}{c}{RTGH} & \multicolumn{1}{c}{Donghae} \\
    \multicolumn{1}{c}{east} & \multicolumn{1}{c}{SUK} & \multicolumn{1}{c}{Samcheok} & \multicolumn{1}{c}{RSCK} & \multicolumn{1}{c}{Samcheok} \\
    \multicolumn{1}{c}{east} & \multicolumn{1}{c}{OKK} & \multicolumn{1}{c}{Okkye} & \multicolumn{1}{c}{RKAG} & \multicolumn{1}{c}{Gangneung} \\
    \multicolumn{1}{c}{east} & \multicolumn{1}{c}{MIP} & \multicolumn{1}{c}{Mipo} & \multicolumn{1}{c}{RUSN} & \multicolumn{1}{c}{Ulsan} \\
    \multicolumn{1}{c}{east} & \multicolumn{1}{c}{ONS} & \multicolumn{1}{c}{Onsan} & \multicolumn{1}{c}{RUSN} & \multicolumn{1}{c}{Ulsan} \\
    \multicolumn{1}{c}{east} & \multicolumn{1}{c}{USN} & \multicolumn{1}{c}{Ulsan} & \multicolumn{1}{c}{RUSN} & \multicolumn{1}{c}{Ulsan} \\
    \multicolumn{1}{c}{east} & \multicolumn{1}{c}{KPO} & \multicolumn{1}{c}{Pohang} & \multicolumn{1}{c}{RPHG} & \multicolumn{1}{c}{Pohang} \\
    \multicolumn{1}{c}{east} & \multicolumn{1}{c}{SHO} & \multicolumn{1}{c}{Sokcho} & \multicolumn{1}{c}{RSHO} & \multicolumn{1}{c}{Sokcho} \\
    \multicolumn{1}{c}{west} & \multicolumn{1}{c}{GIN} & \multicolumn{1}{c}{Gyeong-in} & \multicolumn{1}{c}{RICN} & \multicolumn{1}{c}{Incheon} \\
    \multicolumn{1}{c}{west} & \multicolumn{1}{c}{INC} & \multicolumn{1}{c}{Incheon} & \multicolumn{1}{c}{RICN} & \multicolumn{1}{c}{Incheon} \\
    \multicolumn{1}{c}{west} & \multicolumn{1}{c}{TSN} & \multicolumn{1}{c}{Daesan} & \multicolumn{1}{c}{RSSN} & \multicolumn{1}{c}{Seosan} \\
    \multicolumn{1}{c}{west} & \multicolumn{1}{c}{KUV} & \multicolumn{1}{c}{Gunsan} & \multicolumn{1}{c}{RGSN} & \multicolumn{1}{c}{Gunsan} \\
    \multicolumn{1}{c}{west} & \multicolumn{1}{c}{CHG} & \multicolumn{1}{c}{Janghang} & \multicolumn{1}{c}{RSEO} & \multicolumn{1}{c}{Seocheon$^2$} \\
    \multicolumn{1}{c}{west} & \multicolumn{1}{c}{TJI} & \multicolumn{1}{c}{Dangjin} & \multicolumn{1}{c}{RTJI} & \multicolumn{1}{c}{Dangjin} \\
    \multicolumn{1}{c}{west} & \multicolumn{1}{c}{PTK} & \multicolumn{1}{c}{Pyeongtaek} & \multicolumn{1}{c}{RPTK} & \multicolumn{1}{c}{Pyeongtaek} \\
    \multicolumn{1}{c}{west} & \multicolumn{1}{c}{BOR} & \multicolumn{1}{c}{Boryeong} & \multicolumn{1}{c}{RBOR} & \multicolumn{1}{c}{Boryeong} \\
    \multicolumn{1}{c}{west} & \multicolumn{1}{c}{TAN} & \multicolumn{1}{c}{Taean} & \multicolumn{1}{c}{RTAN} & \multicolumn{1}{c}{Taean$^3$} \\
    \midrule
    \multicolumn{5}{l}{\textit{$^{1,2,3}$  The level of these port regions is a county, called ’gun’ in Korean.}} \\
    \end{tabular*}%
  \label{tab:the sample}%
  }
\end{table}%

\begin{table}[htb!]
  \centering
  \caption{The containerization ratio by item, the case of Korea in 2020}
    \begin{tabular}{ll}
    \toprule
    \multicolumn{1}{c}{\textbf{Item}} & \multicolumn{1}{c}{\textbf{Containerization ratios }} \\
    \midrule
    \multicolumn{1}{c}{Meat and edible meat offal} & \multicolumn{1}{c}{0.99 } \\
    \multicolumn{1}{c}{Fish and other aquatic invertebrates} & \multicolumn{1}{c}{0.86 } \\
    \multicolumn{1}{c}{Cereals} & \multicolumn{1}{c}{0.13 } \\
    \multicolumn{1}{c}{Malt, starches, inulin, wheat gluten} & \multicolumn{1}{c}{0.56 } \\
    \multicolumn{1}{c}{Animals; live} & \multicolumn{1}{c}{0.85 } \\
    \multicolumn{1}{c}{Animal or vegetable fats and oils } & \multicolumn{1}{c}{0.48 } \\
    \multicolumn{1}{c}{Sugars and sugar confectionery} & \multicolumn{1}{c}{0.42 } \\
    \multicolumn{1}{c}{Meat, fish, or preparations thereof} & \multicolumn{1}{c}{0.72 } \\
    \multicolumn{1}{c}{Cement} & \multicolumn{1}{c}{0.01 } \\
    \multicolumn{1}{c}{Sands and earths} & \multicolumn{1}{c}{0.16 } \\
    \multicolumn{1}{c}{Hard coal} & \multicolumn{1}{c}{0.01 } \\
    \multicolumn{1}{c}{Bituminous coal} & \multicolumn{1}{c}{0.00 } \\
    \multicolumn{1}{c}{Ores, slag and ash} & \multicolumn{1}{c}{0.00 } \\
    \multicolumn{1}{c}{Salt, sulphur and others} & \multicolumn{1}{c}{0.19 } \\
    \multicolumn{1}{c}{Crude oil and bituminous substances} & \multicolumn{1}{c}{0.00 } \\
    \multicolumn{1}{c}{Petroleum refining } & \multicolumn{1}{c}{0.01 } \\
    \multicolumn{1}{c}{Petroleum products of their distillation} & \multicolumn{1}{c}{0.00 } \\
    \multicolumn{1}{c}{Fertilizers} & \multicolumn{1}{c}{0.34 } \\
    \multicolumn{1}{c}{Inorganic chemicals and  articles thereof} & \multicolumn{1}{c}{0.48 } \\
    \multicolumn{1}{c}{Plastics and articles thereof} & \multicolumn{1}{c}{1.00 } \\
    \multicolumn{1}{c}{Raw hides, skins, and leather} & \multicolumn{1}{c}{1.00 } \\
    \multicolumn{1}{c}{Wood and articles of wood} & \multicolumn{1}{c}{0.53 } \\
    \multicolumn{1}{c}{Wood  charcoal} & \multicolumn{1}{c}{0.87 } \\
    \multicolumn{1}{c}{Silk} & \multicolumn{1}{c}{1.00 } \\
    \multicolumn{1}{c}{Scrap metal} & \multicolumn{1}{c}{0.26 } \\
    \multicolumn{1}{c}{Iron and steel} & \multicolumn{1}{c}{0.35 } \\
    \multicolumn{1}{c}{Copper and articles thereof} & \multicolumn{1}{c}{0.95 } \\
    \multicolumn{1}{c}{Nuclear reactors, machinery and parts thereof} & \multicolumn{1}{c}{0.90 } \\
    \multicolumn{1}{c}{\textbf{Electrical machinery and parts thereof}} & \multicolumn{1}{c}{0.99 } \\
    \multicolumn{1}{c}{Railway, locomotives, and parts thereof} & \multicolumn{1}{c}{0.34 } \\
    \multicolumn{1}{c}{Aircraft, spacecraft and parts thereof} & \multicolumn{1}{c}{0.95 } \\
    \multicolumn{1}{c}{ETC.} & \multicolumn{1}{c}{0.97 } \\
    \midrule
    \multicolumn{2}{p{37em}}{\textit{$^1$Source : Author's elaboration, using the data of the year 2020, from The Korean Port Management Information System (PORT-MIS)}} \\
    \multicolumn{2}{p{37em}}{\textit{$^2$ The containerization ratio by item is calculated as the ratio of containerized cargo to the total export volume by 28 major ports.}} \\
    \end{tabular}%
  \label{tab:cr}%
\end{table}%

\begin{table}[hbt!]
  \centering
  \caption{Summary statistics of variables for testing (a) inter-industry spillover from related industries within a region and (b) cross-space spillover from neighboring ports to port regions }
    \begin{tabular}{c p{5em} m{5em} m{4em} m{4em} m{4em} m{4em}}
    \toprule
          \multicolumn{1}{l}{} & \multicolumn{1}{l}{\textbf{Variable}} & \multicolumn{1}{l}{\textbf{Observations}} & \multicolumn{1}{l}{\textbf{Mean}} & \multicolumn{1}{l}{\textbf{Std. Dev.}} & \multicolumn{1}{l}{\textbf{Min}} & \multicolumn{1}{l}{\textbf{Max}} \\
    \midrule
    \multicolumn{1}{c}{\multirow{4}[2]{*}{(a)}} &  $\omega_{r,i}^t$ & 33,057  & 0.234 & 0.229 & 0.001 & 0.903 \\
          &  $k_{r,i}^t$ & 33,057  & 25.0  & 14.2  & 1     & 72 \\
          &  $TRM_{r,i}^t$ & 33,057  & 1.4   & 0.8   & 1     & 9 \\
          &  $PCI_{i}^t$  & 33,057  & 0.240  & 0.934 & -3.273 & 2.644 \\
    \midrule
    \multicolumn{1}{c}{\multirow{6}[2]{*}{(b)}} &  $\omega_{r,i}^t$ & 13,744  & 0.346 & 0.266 & 0.002 & 0.903 \\
          &  $\Omega_{p,i}^t$ & 13,744  & 0.419 & 0.331 & 0     & 0.977 \\
          &  $k_{p,i}^t$ & 13,744  & 20.9  & 13.6  & 1     & 72 \\
          &  $k_{r,i}^t$ & 13,744  & 4.9   & 1.9   & 1     & 15 \\
          &  $TRM_{r,i}^t$ & 13,744  & 1.7   & 1.1   & 1     & 9 \\
          &  $PCI_{i}^t$  & 13,744  & 0.278 & 0.917 & -3.273 & 2.526 \\
    \bottomrule
    \end{tabular}
  \label{tab:statistics}%
\end{table}%

\begin{table}[htb!]
  \centering
  \caption{Robustness check for Eq.\ref{eq:model1}}
     \resizebox{\textwidth}{!}{\begin{tabular}{lllllllll}
    \toprule
    \multicolumn{1}{c}{   } & \multicolumn{1}{c}{(1)} & \multicolumn{1}{c}{(2)} & \multicolumn{1}{c}{(3)} & \multicolumn{1}{c}{(4)} &       & \multicolumn{1}{c}{(5)} & \multicolumn{1}{c}{(6)} & \multicolumn{1}{c}{(7)} \\
\cmidrule{2-5}\cmidrule{7-9}    \multicolumn{1}{c}{$S_{r,i}^{t+2}$} & \multicolumn{4}{c}{\makecell{Probit, \\applying variables one by one}} &       & \multicolumn{1}{c}{Probit} & \multicolumn{1}{c}{Logit} & \multicolumn{1}{c}{LPM} \\
\cmidrule{1-5}\cmidrule{7-9}    \multicolumn{1}{c}{$\omega_{r,i}^t$} & \multicolumn{1}{c}{7.230***} &       &       &       &       & \multicolumn{1}{c}{7.041***} & \multicolumn{1}{c}{12.044***} & \multicolumn{1}{c}{2.077***} \\
    \multicolumn{1}{c}{  } & \multicolumn{1}{c}{(0.315)} &       &       &       &       & \multicolumn{1}{c}{(0.342)} & \multicolumn{1}{c}{(0.615)} & \multicolumn{1}{c}{(0.076)} \\
    \multicolumn{1}{c}{$k_{r,i}^t$} &       & \multicolumn{1}{c}{0.015***} &       &       &       & \multicolumn{1}{c}{0.007***} & \multicolumn{1}{c}{0.013***} & \multicolumn{1}{c}{0.001***} \\
    \multicolumn{1}{c}{  } &       & \multicolumn{1}{c}{(0.001)} &       &       &       & \multicolumn{1}{c}{(0.002)} & \multicolumn{1}{c}{(0.003)} & \multicolumn{1}{c}{0.000 } \\
    \multicolumn{1}{c}{$PCI_{i}^t$} &       &       & \multicolumn{1}{c}{-0.074***} &       &       & \multicolumn{1}{c}{-0.058***} & \multicolumn{1}{c}{-0.087**} & \multicolumn{1}{c}{-0.011**} \\
    \multicolumn{1}{c}{  } &       &       & \multicolumn{1}{c}{(0.021)} &       &       & \multicolumn{1}{c}{(0.019)} & \multicolumn{1}{c}{(0.034)} & \multicolumn{1}{c}{(0.005)} \\
    \multicolumn{1}{c}{$TRM_{r,i}^t$} &       &       &       & \multicolumn{1}{c}{0.056**} &       & \multicolumn{1}{c}{0.156***} & \multicolumn{1}{c}{0.272***} & \multicolumn{1}{c}{0.041***} \\
    \multicolumn{1}{c}{  } &       &       &       & \multicolumn{1}{c}{(0.026)} &       & \multicolumn{1}{c}{(0.021)} & \multicolumn{1}{c}{(0.036)} & \multicolumn{1}{c}{(0.006)} \\
    \multicolumn{1}{c}{$Constant$} & \multicolumn{1}{c}{-1.722***} & \multicolumn{1}{c}{-2.247***} & \multicolumn{1}{c}{-1.747***} & \multicolumn{1}{c}{-1.842***} &       & \multicolumn{1}{c}{-2.135***} & \multicolumn{1}{c}{-3.727***} & \multicolumn{1}{c}{-0.081***} \\
    \multicolumn{1}{c}{  } & \multicolumn{1}{c}{(0.098)} & \multicolumn{1}{c}{(0.108)} & \multicolumn{1}{c}{(0.098)} & \multicolumn{1}{c}{(0.100)} &       & \multicolumn{1}{c}{(0.105)} & \multicolumn{1}{c}{(0.197)} & \multicolumn{1}{c}{(0.019)} \\
    \multicolumn{1}{c}{Year dummies} & \multicolumn{1}{c}{yes} & \multicolumn{1}{c}{yes} & \multicolumn{1}{c}{yes} & \multicolumn{1}{c}{yes} &       & \multicolumn{1}{c}{yes} & \multicolumn{1}{c}{yes} & \multicolumn{1}{c}{yes} \\
    \multicolumn{1}{c}{Region dummies} & \multicolumn{1}{c}{yes} & \multicolumn{1}{c}{ yes} & \multicolumn{1}{c}{ yes} & \multicolumn{1}{c}{ yes} &       & \multicolumn{1}{c}{ yes} & \multicolumn{1}{c}{ yes} & \multicolumn{1}{c}{ yes} \\
          &       &       &       &       &       &       &       &  \\
    \multicolumn{1}{c}{Observations} & \multicolumn{1}{c}{33,057 } & \multicolumn{1}{c}{33,057 } & \multicolumn{1}{c}{33,057 } & \multicolumn{1}{c}{33,057 } &       & \multicolumn{1}{c}{33,057 } & \multicolumn{1}{c}{33,057 } & \multicolumn{1}{c}{33,057 } \\
    \multicolumn{1}{c}{Pseudo R$^2$} & \multicolumn{1}{c}{0.125 } & \multicolumn{1}{c}{0.099 } & \multicolumn{1}{c}{0.085 } & \multicolumn{1}{c}{0.084 } &       & \multicolumn{1}{c}{0.136 } & \multicolumn{1}{c}{0.135 } & \multicolumn{1}{c}{0.142\^3} \\
    \multicolumn{1}{c}{log-likelihood} & \multicolumn{1}{c}{-14421 } & \multicolumn{1}{c}{-14860 } & \multicolumn{1}{c}{-15087 } & \multicolumn{1}{c}{-15105 } &       & \multicolumn{1}{c}{-14246 } & \multicolumn{1}{c}{-14256 } & \multicolumn{1}{c}{-14018 } \\
    \multicolumn{1}{c}{Mean VIF} & \multicolumn{1}{c}{4.70 } & \multicolumn{1}{c}{2.00 } & \multicolumn{1}{c}{1.85 } & \multicolumn{1}{c}{1.97 } &       & \multicolumn{1}{c}{5.24 } & \multicolumn{1}{c}{5.24 } & \multicolumn{1}{c}{7.87 } \\
    \midrule
    \multicolumn{9}{l}{\textit{$^1$ Standard errors are in parentheses}} \\
    \multicolumn{9}{l}{\textit{$^2$*** p$<$ 0.01, ** p$<$ 0.05, * p$<$ 0.1 }}\\
    \multicolumn{9}{l}{\textit{$^3$ This is the value of R$^2$ in the case of a linear probability model(LPM).}} \\
    \multicolumn{9}{p{45em}}{\textit{$^4$ Mean VIF represents the mean of uncentered variance inflation factors (VIFs) for all  applied variables, detecting the collinearity.}} \\
    \end{tabular}}%
  \label{tab:robust1}%
\end{table}%

\begin{landscape}
\begin{table}[htb!]
  \centering
  \caption{Robustness check for Eq.\ref{eq:model2}}
    \begin{tabular}{ccccccccccc}
    \toprule
          & \multicolumn{1}{c}{(1)} & \multicolumn{1}{c}{(2)} & \multicolumn{1}{c}{(3)} & \multicolumn{1}{c}{(4)} & \multicolumn{1}{c}{(5)} & \multicolumn{1}{c}{(6)} &       & \multicolumn{1}{c}{(7)} & \multicolumn{1}{c}{(8)} & \multicolumn{1}{c}{(9)} \\
\cmidrule{2-7}\cmidrule{9-11}    $S_{r,p,i}^{t+2}$ & \multicolumn{6}{c}{\makecell{Probit, \\applying variables one by one}} &       & \multicolumn{1}{c}{Probit} & \multicolumn{1}{c}{Logit} & \multicolumn{1}{c}{LPM} \\
\cmidrule{1-7}\cmidrule{9-11}    $\omega_{r,i}^t$ & \multicolumn{1}{c}{7.818***} &       &       &       &       &       &       & \multicolumn{1}{c}{7.833***} & \multicolumn{1}{c}{13.414***} & \multicolumn{1}{c}{2.329***} \\
          & \multicolumn{1}{c}{(0.384)} &       &       &       &       &       &       & \multicolumn{1}{c}{(0.424)} & \multicolumn{1}{c}{(0.775)} & \multicolumn{1}{c}{(0.090)} \\
    $\Omega_{p,i}^t$ &       & \multicolumn{1}{c}{1.370***} &       &       &       &       &       & \multicolumn{1}{c}{0.707***} & \multicolumn{1}{c}{1.248***} & \multicolumn{1}{c}{0.194***} \\
    \multicolumn{1}{l}{} &       & \multicolumn{1}{c}{(0.233)} &       &       &       &       &       & \multicolumn{1}{c}{(0.175)} & \multicolumn{1}{c}{(0.302)} & \multicolumn{1}{c}{(0.044)} \\
    $k_{r,i}^t$    &       &       & \multicolumn{1}{c}{0.056***} &       &       &       &       & \multicolumn{1}{c}{0.056***} & \multicolumn{1}{c}{0.097***} & \multicolumn{1}{c}{0.014***} \\
    \multicolumn{1}{l}{} &       &       & \multicolumn{1}{c}{(0.015)} &       &       &       &       & \multicolumn{1}{c}{(0.012)} & \multicolumn{1}{c}{(0.020)} & \multicolumn{1}{c}{(0.003)} \\
    $K_{p,i}^t$     &       &       &       & \multicolumn{1}{c}{0.024***} &       &       &       & \multicolumn{1}{c}{0.006***} & \multicolumn{1}{c}{0.011***} & \multicolumn{1}{c}{0.001**} \\
          &       &       &       & \multicolumn{1}{c}{(0.002)} &       &       &       & \multicolumn{1}{c}{(0.002)} & \multicolumn{1}{c}{(0.004)} & \multicolumn{1}{c}{0.000 } \\
    $PCI_{i}^t$   &       &       &       &       & \multicolumn{1}{c}{-0.054*} &       &       & \multicolumn{1}{c}{-0.010 } & \multicolumn{1}{c}{0.000 } & \multicolumn{1}{c}{0.001} \\
          &       &       &       &       & \multicolumn{1}{c}{(0.028)} &       &       & \multicolumn{1}{c}{(0.025)} & \multicolumn{1}{c}{(0.043)} & \multicolumn{1}{c}{(0.006)} \\
    $TRM_{r,i}^t$   &       &       &       &       &       & \multicolumn{1}{c}{-0.014 } &       & \multicolumn{1}{c}{0.111***} & \multicolumn{1}{c}{0.197***} & \multicolumn{1}{c}{0.029***} \\
          &       &       &       &       &       & \multicolumn{1}{c}{(0.026)} &       & \multicolumn{1}{c}{(0.022)} & \multicolumn{1}{c}{(0.036)} & \multicolumn{1}{c}{(0.006)} \\
    $Constant$ & \multicolumn{1}{c}{-2.349***} & \multicolumn{1}{c}{-1.856***} & \multicolumn{1}{c}{-2.164***} & \multicolumn{1}{c}{-2.330***} & \multicolumn{1}{c}{-1.798***} & \multicolumn{1}{c}{-1.805***} &       & \multicolumn{1}{c}{-3.001***} & \multicolumn{1}{c}{-5.197***} & \multicolumn{1}{c}{-0.239***} \\
          & \multicolumn{1}{c}{(0.237)} & \multicolumn{1}{c}{(0.230)} & \multicolumn{1}{c}{(0.248)} & \multicolumn{1}{c}{(0.240)} & \multicolumn{1}{c}{(0.230)} & \multicolumn{1}{c}{(0.232)} &       & \multicolumn{1}{c}{(0.249)} & \multicolumn{1}{c}{(0.469)} & \multicolumn{1}{c}{(0.029)} \\
    Year dummies & \multicolumn{1}{c}{yes} & \multicolumn{1}{c}{yes} & \multicolumn{1}{c}{yes} & \multicolumn{1}{c}{yes} & \multicolumn{1}{c}{yes} & \multicolumn{1}{c}{yes} &       & \multicolumn{1}{c}{yes} & \multicolumn{1}{c}{yes} & \multicolumn{1}{c}{yes} \\
    Region dummies & \multicolumn{1}{c}{yes} & \multicolumn{1}{c}{ yes} & \multicolumn{1}{c}{ yes} & \multicolumn{1}{c}{ yes} & \multicolumn{1}{c}{ yes} & \multicolumn{1}{c}{ yes} &       & \multicolumn{1}{c}{ yes} & \multicolumn{1}{c}{ yes} & \multicolumn{1}{c}{ yes} \\
    \multicolumn{1}{r}{} &       &       &       &       &       &       &       &       &       &  \\
    Observations & \multicolumn{1}{c}{13,735 } & \multicolumn{1}{c}{13,735 } & \multicolumn{1}{c}{13,735 } & \multicolumn{1}{c}{13,735 } & \multicolumn{1}{c}{13,735 } & \multicolumn{1}{c}{13,735 } &       & \multicolumn{1}{c}{13,735 } & \multicolumn{1}{c}{13,735 } & \multicolumn{1}{c}{13,744 } \\
    Pseudo R$^2$ & \multicolumn{1}{c}{0.193 } & \multicolumn{1}{c}{0.119 } & \multicolumn{1}{c}{0.116 } & \multicolumn{1}{c}{0.142 } & \multicolumn{1}{c}{0.113 } & \multicolumn{1}{c}{0.113 } &       & \multicolumn{1}{c}{0.206 } & \multicolumn{1}{c}{0.206 } & \multicolumn{1}{c}{0.222 } \\
    log-likelihood & \multicolumn{1}{c}{-6241 } & \multicolumn{1}{c}{-6813 } & \multicolumn{1}{c}{-6830 } & \multicolumn{1}{c}{-6629 } & \multicolumn{1}{c}{-6853 } & \multicolumn{1}{c}{-6860 } &       & \multicolumn{1}{c}{-6138 } & \multicolumn{1}{c}{-6139 } & \multicolumn{1}{c}{-6276 } \\
    Mean VIF & \multicolumn{1}{c}{5.42 } & \multicolumn{1}{c}{4.55 } & \multicolumn{1}{c}{2.67 } & \multicolumn{1}{c}{2.44 } & \multicolumn{1}{c}{2.28 } & \multicolumn{1}{c}{2.41 } &       & \multicolumn{1}{c}{8.33 } & \multicolumn{1}{c}{8.33 } & \multicolumn{1}{c}{663.32 } \\
    \midrule
    \multicolumn{11}{l}{\textit{$^1$ Standard errors are in parentheses}} \\
    \multicolumn{11}{l}{\textit{$^2$*** p$<$ 0.01, ** p$<$ 0.05, * p$<$ 0.1 }}\\
    \multicolumn{11}{l}{\textit{$^3$ This is the value of R$^2$ in the case of a linear probability model(LPM).}} \\
    \multicolumn{11}{p{55em}}{\textit{$^4$ Mean VIF represents the mean of uncentered variance inflation factors (VIFs) for all applied variables, detecting the collinearity.}} \\
    \end{tabular}%
  \label{tab:robust2}%
\end{table}%
\end{landscape}

\begin{table}[htb!]
  \centering
  \caption{The development of a new industry in times of economic crisis (2007-2009), recovery (2010-2013) and post-crisis (after 2014)}
    \resizebox{\textwidth}{!}{\begin{tabular}{c m{10em} m{10em} m{10em}}
    \toprule
          & \multicolumn{1}{c}{(1)} & \multicolumn{1}{c}{(2)} & \multicolumn{1}{c}{(3)} \\
\cmidrule{2-4}    $S_{r,p,i}^{t+2}$ & \multicolumn{1}{c}{\makecell{Economic crisis\\(2007-2009)}} & \multicolumn{1}{c}{\makecell{Recovery\\(2010-2013)}} & \multicolumn{1}{c}{\makecell{Post-crisis\\(2014-2018)}  } \\
    \midrule
    $\omega_{r,i}^t$ & \multicolumn{1}{c}{6.903***} & \multicolumn{1}{c}{8.405***} & \multicolumn{1}{c}{7.243***} \\
          & \multicolumn{1}{c}{(1.073)} & \multicolumn{1}{c}{(0.673)} & \multicolumn{1}{c}{(0.570)} \\
    $\Omega_{p,i}^t$ & \multicolumn{1}{c}{0.512} & \multicolumn{1}{c}{0.728*} & \multicolumn{1}{c}{0.547***} \\
    \multicolumn{1}{l}{} & \multicolumn{1}{c}{(0.619)} & \multicolumn{1}{c}{(0.420)} & \multicolumn{1}{c}{(0.199)} \\
    $k_{r,i}^t$     & \multicolumn{1}{c}{0.009**} & \multicolumn{1}{c}{0.003} & \multicolumn{1}{c}{0.003} \\
    \multicolumn{1}{l}{} & \multicolumn{1}{c}{(0.004)} & \multicolumn{1}{c}{(0.003)} & \multicolumn{1}{c}{(0.003)} \\
    $K_{p,i}^t$     & \multicolumn{1}{c}{0.027} & \multicolumn{1}{c}{0.062***} & \multicolumn{1}{c}{0.091***} \\
          & \multicolumn{1}{c}{(0.021)} & \multicolumn{1}{c}{(0.018)} & \multicolumn{1}{c}{(0.017)} \\
    $PCI_{i}^t$  & \multicolumn{1}{c}{0.055 } & \multicolumn{1}{c}{0.030 } & \multicolumn{1}{c}{-0.013 } \\
          & \multicolumn{1}{c}{(0.045)} & \multicolumn{1}{c}{(0.032)} & \multicolumn{1}{c}{(0.032)} \\
    $DES_{p,i}^t$   & \multicolumn{1}{c}{0.017} & \multicolumn{1}{c}{-0.015} & \multicolumn{1}{c}{-0.036**} \\
          & \multicolumn{1}{c}{(0.025)} & \multicolumn{1}{c}{(0.019)} & \multicolumn{1}{c}{(0.017)} \\
    $Constant$ & \multicolumn{1}{c}{-7.027***} & \multicolumn{1}{c}{-8.453***} & \multicolumn{1}{c}{-2.985***} \\
          & \multicolumn{1}{c}{(1.008)} & \multicolumn{1}{c}{(0.689)} & \multicolumn{1}{c}{(0.301)} \\
    Year dummies & \multicolumn{1}{c}{yes} & \multicolumn{1}{c}{yes} & \multicolumn{1}{c}{yes} \\
    Region dummies & \multicolumn{1}{c}{ yes} & \multicolumn{1}{c}{ yes} & \multicolumn{1}{c}{ yes} \\
    \multicolumn{1}{r}{} &       &       &  \\
    Observations & \multicolumn{1}{c}{2,820 } & \multicolumn{1}{c}{4,748 } & \multicolumn{1}{c}{5,450 } \\
    Pseudo R$^2$ & \multicolumn{1}{c}{0.188} & \multicolumn{1}{c}{0.203} & \multicolumn{1}{c}{0.213} \\
    log-likelihood & \multicolumn{1}{c}{-1269.959} & \multicolumn{1}{c}{-2115.669} & \multicolumn{1}{c}{-2543.948} \\
    Mean VIF & \multicolumn{1}{c}{7.50 } & \multicolumn{1}{c}{5.37 } & \multicolumn{1}{c}{9.29 } \\
     \midrule
    \multicolumn{4}{p{40.26em}}{\textit{$^1$Standard errors are in parentheses}} \\
    \multicolumn{4}{p{40.26em}}{\textit{$^2$*** p$<$ 0.01, ** p$<$ 0.05, * p$<$ 0.1 }} \\
    \multicolumn{4}{p{40.26em}}{\textit{$^3$ Mean VIF represents the mean of uncentered variance inflation factors (VIFs) for all applied variables, detecting the collinearity.}} \\
    \end{tabular}}%
  \label{tab:postcrisis}%
\end{table}%

\begin{figure}[htb!]
\centering
\includegraphics[width=1\textwidth]{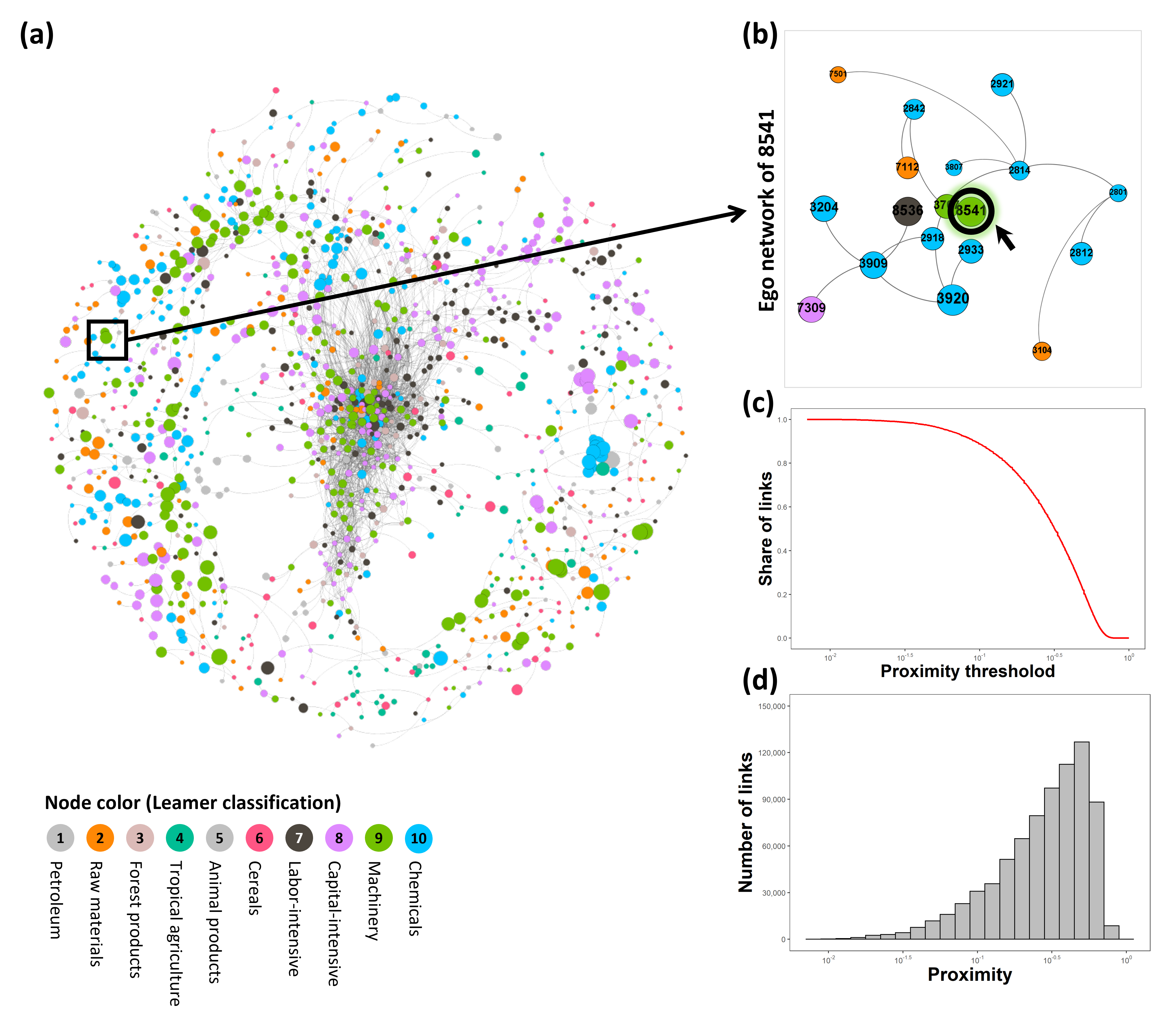}
  \caption{Product space based on the production proximity $\phi$, 2006-2020 : (a) Network representation of all products, (b) Ego network of the product of HS code 8541 (depth=3), (c) cumulative distribution of the production proximity $\phi$, and (d) density distribution of $\phi$ \\ $^1$\textit{Product spaces are built following \cite{Hidalgo2007} and \cite{Gao2021}.} $^2$ \textit{The node size represents the export value of the product in a relative scale, and the node color shows its classification as originally proposed by \cite{Leamer1984} and revised by \cite{Hidalgo2007}}}
  \label{fig:r_ps}
\end{figure}

\begin{figure}[htb!]
\centering
\includegraphics[width=1\textwidth]{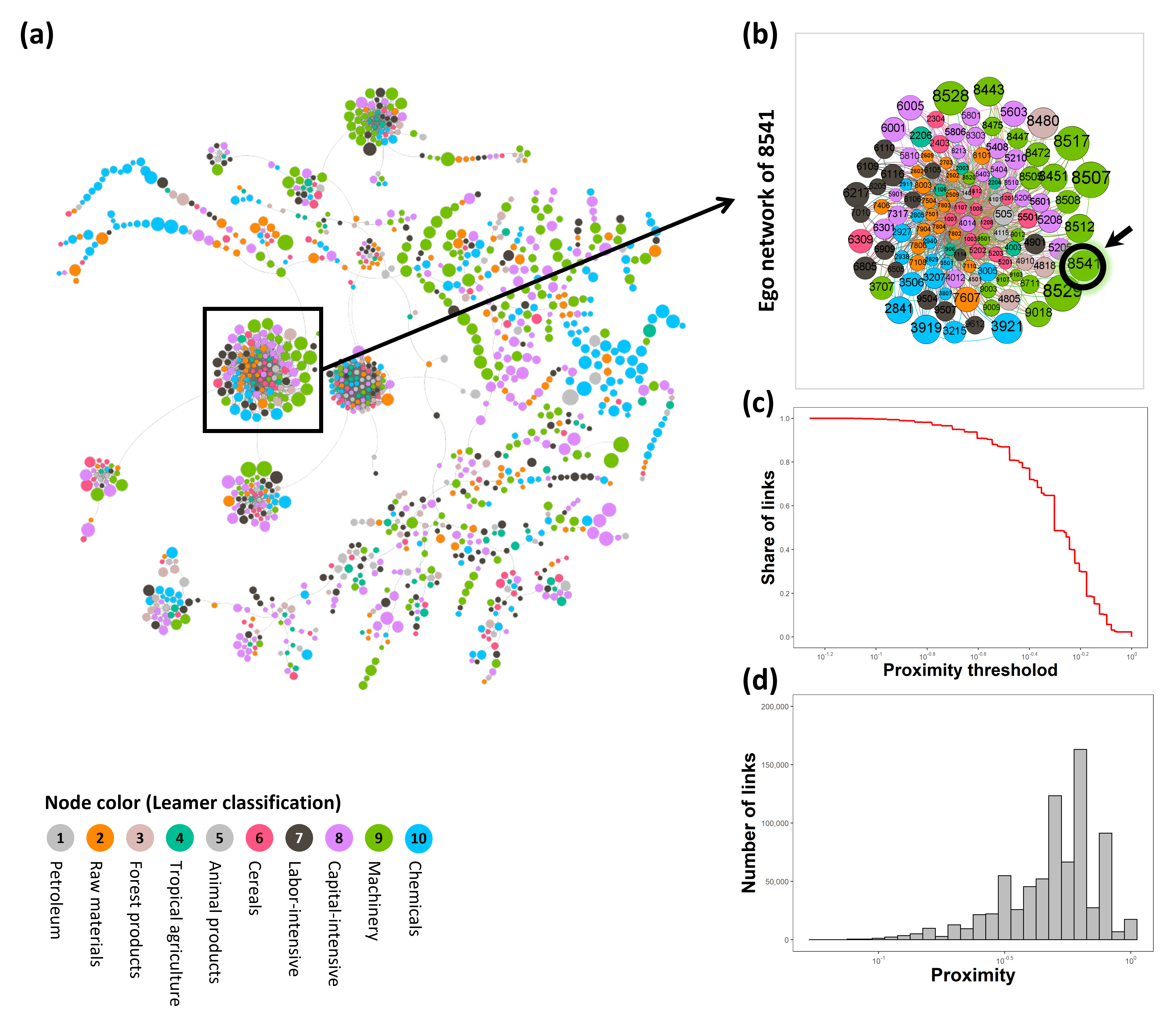}
  \caption{Product space based on the transport proximity $\Phi$, 2006-2020 : (a) Network representation of all products, (b) Ego network of the product of HS code 8541 (depth=3), (c) cumulative distribution of the production proximity $\Phi$, and (d) density distribution of $\Phi$ \\ $^1$\textit{Product spaces are built following \cite{Hidalgo2007} and \cite{Gao2021}.} $^2$ \textit{The node size represents the export value of the product in a relative scale, and the node color shows its classification as originally proposed by \cite{Leamer1984} and revised by \cite{Hidalgo2007}}}
  \label{fig:p_ps}
\end{figure}


\end{document}